\documentclass[]{aastex63}
\usepackage{amsmath}
\usepackage[caption=false]{subfig}
\shorttitle{Water in VV CrA}
\shortauthors{Salyk et al.}

\begin{document}
\title{An unusual reservoir of water emission in the VV CrA A protoplanetary disk}

\correspondingauthor{Colette Salyk}
\email{cosalyk@vassar.edu}

\author[0000-0003-3682-6632]{Colette Salyk}
\affiliation{Vassar College \\
124 Raymond Avenue\\
Poughkeepsie, NY 12604, USA}

\author[0000-0002-0786-7307]{Klaus M. Pontoppidan}
\affil{Space Telescope Science Institute \\
3700 San Martin Drive \\
Baltimore, MD 21218, USA}

\author[0000-0003-4335-0900]{Andrea Banzatti}
\affil{Department of Physics\\
Texas State University\\
749 N Comanche St\\
San Marcos, TX 78666, USA}

\author[0000-0002-8806-9795]{Ulrich K{\"a}ufl}
\affil{Josef-Raps-Sttrasse 5\\
D-80805, M\:{u}nchen, Germany}

\author[0000-0002-8138-0425]{Cassandra Hall}
\affil{Department of Physics and Astronomy\\
The University of Georgia\\
Athens, GA 30602, USA}
\affil{Center for Simulational Physics\\
The University of Georgia\\
Athens, GA 30602, USA}
\affil{School of Physics and Astronomy\\
University of Leicester\\
Leicester, LE1 7RH, UK}

\author[0000-0001-7962-1683]{Ilaria Pascucci}
\affil{Lunar and Planetary Laboratory\\
The University of Arizona\\
Tucson, AZ 85721, USA}

\author[0000-0003-2471-1299]{Andr{\'e}s Carmona}
\affil{Universit\'e Grenoble Alpes\\
 IPAG\\
 38000, Grenoble, France}
\affil{CNRS, IPAG
38000, Grenoble, France}
 
\author[0000-0003-0787-1610]{Geoffrey A. Blake}
\affil{Division of Geological and Planetary Sciences\\
California Institute of Technology\\
Pasadena, CA 91125, USA}

\author[0000-0001-6410-2899]{Richard Alexander}
\affil{School of Physics and Astronomy\\ 
University of Leicester\\
Leicester, LE1 7RH, UK}

\author[0000-0001-6410-2899]{Inga Kamp}
\affil{Kapteyn Institute\\
PO Box 800\\
9700 AV Groningen\\
The Netherlands}

\begin{abstract}
We present an analysis of an unusual pattern of water vapor emission from the $\sim$2 Myr-old low-mass binary system VV CrA, as observed in infrared spectra obtained with VLT-CRIRES, VLT-VISIR, and Spitzer-IRS. Each component of the binary shows emission from water vapor in both the L ($\sim3\,\mu$m) and N ($\sim 12\,\mu$m) bands. The N-band and Spitzer spectra are similar to those previously observed from young stars with disks, and are consistent with emission from an extended protoplanetary disk. Conversely, the CRIRES L-band data of VV CrA A show an unusual spectrum, which requires the presence of a water reservoir with high temperature ($T\gtrsim1500$ K), column density ($N_\mathrm{H2O}\sim 3\times10^{20}\ \mathrm{cm}^{-2}$), and turbulent broadening ($v\sim 10$ km s$^{-1}$), but very small emitting area ($A\lesssim0.005$ AU$^2$).  Similarity with previously observed water emission from V1331 Cyg \citep{Doppmann11} and SVS 13 \citep{Carr04} suggests that the presence of such a reservoir may be linked to evolutionary state, perhaps related to the presence of high accretion rates or winds. While the inner disk may harbor such a reservoir, simple Keplerian models do not match well with emitting line shapes, and alternative velocity fields must be considered.  We also present a new idea, that the unusual emission could arise in a circumplanetary disk, embedded within the larger VV CrA A protoplanetary disk.  Additional data are likely required to determine the true physical origin of this unusual spectral pattern.

\end{abstract}

\section{Introduction}
The properties of planets are set by the physical and chemical conditions within the protoplanetary disks from which they form. Local disk chemistry may determine not only the bulk elemental composition of a planet \citep[e.g.][]{Grossman72}, but may also have profound effects on a wide range of planet properties. Whether or not ice is in solid form may drive planetesimal formation \citep{Drazkowska17}, or increase surface densities sufficiently to allow for gas giant formation \citep[e.g.,][]{Stevenson88, Kennedy08}. Interior and atmospheric chemistry can influence a planet's thermal evolution \citep[e.g.,][]{Head81}, surface temperature \citep[e.g.,][]{Kopparapu13}, and ultimately, its suitability for life. However, it may not be possible to use ab initio chemico-dynamical models to reliably predict the planetary system that results from a given set of initial conditions. As recent ALMA observations of rings, gaps, and spirals in disks reveal \citep[e.g.][]{Andrews18}, protoplanets and planets may already be present in intermediate age ``Class II'' disks and are, themselves, influencing the disk environment. Thus, there is a mutual co-evolution of both the planet and the disk structure which will, in turn, affect the chemistry, potentially in chaotic ways \citep[e.g.][]{Salyk15}. Therefore, ground-truth observations of disk chemistry are required to establish how disk chemistry begins, and how it evolves.

In particular, inner disk ($R\lesssim 10\,$AU) chemistry can be studied with gas-phase infrared spectroscopy, which probes the warm, molecular layer at relatively high disk altitudes \citep{Najita03}. To date, the Spitzer InfraRed Spectrograph (IRS) has provided the majority of molecular detections \citep[e.g.][]{Carr08,Pontoppidan10a}. This includes the detection of water vapor, a major molecular constituent of disks whose gas phase abundance is influenced by such factors as the local radiation field and gas density \citep{Du14}, and the radial transport of icy solids \citep{Ciesla06}. However, Spitzer-IRS data are both spatially and spectrally unresolved, such that they reveal no kinematic information. Ground-based high resolution spectroscopy is a powerful supplementary tool for locating the water and other molecules within the disk. 

Much of the ground-based spectroscopic study of water vapor in disks \citep[e.g.][]{Banzatti17} has been performed with the Very Large Telescope (VLT)'s CRyogenic high-resolution InfraRed Echelle Spectrograph (CRIRES; \citealp{Kaeufl04}). Mid-infrared water emission has been studied previously with Michelle \citep{Glasse97} and TEXES \citep{Lacy02}, confirming that the mid-IR emission lines arise from disk surfaces within the water snow line \citep{Salyk19}.  More recently, we have executed a large survey of water vapor in the mid-infrared (N-band) regime with the VLT Imager and Spectrometer for the mid-InfraRed (VISIR; \citealp{Lagage04,Kaufl15}).

One avenue of disk chemistry that remains relatively unexplored is chemistry in binary (or multiple) systems, and to understand the interaction between disk chemistry and companions, including protoplanets. To what extent does the presence of companions affect chemistry? Do disks in multiple systems, which are both coeval and derive from the same initial conditions, develop the same disk chemistry? The latter question provides a powerful test of the extent to which disk (and therefore planetary) chemistry is deterministic or stochastic. It was with these questions in mind that we added the binary VV CrA to our spectroscopic survey. 
VV CrA is a $\sim$ 2 Myr-old \citep{Sullivan19} 2.1$''$ binary located in the Corona Australis star forming region, at a distance of 157 pc (GAIA DR3; \citealp{Gaia16,Gaia21}). It is associated with a Herbig-Haro object \citep{Wang04}.  The SW component is of type M1 and dominates at visible wavelengths.  The NE component, of type M0, is faint at visible/near-IR wavelengths, but has dominated in the thermal infrared at some epochs \citep{Avilez17}, and has therefore in some references been classified as an InfraRed Companion \citep[IRC,][]{Koresko97}. The photospheric spectroscopy and modeling of \cite{Sullivan19} suggest that the SW and NE stellar components have masses of $\sim 0.55$ and $0.45\,M_{\odot}$, respectively. Following the conventions of previous authors, we will refer to the SW component as the primary (VV CrA A), and the NE component as the secondary infrared companion (VV CrA B). VV CrA A is inclined by 32$^\circ$ \citep{Gravity21} but the geometry of the two disks relative to each other is still under debate; quantitative analyses of absorption features from CO \citep{Smith09} and silicates \citep{Kruger11} suggest that the light from the secondary may be absorbed by the disk of the primary. 

In this paper, we analyze multi-instrument, ground-based high-resolution ($\lambda/\Delta\lambda=R\sim30,000-95,000$) and Spitzer-IRS Short- and Long-High ($R\sim 500$) infrared spectra of an unusual pattern of water vapor emission from the protoplanetary disks in the VV CrA system. Section \ref{sec:observations} presents spectra from VLT-CRIRES, VLT-VISIR, and Spitzer-IRS, covering wavelengths from 2.3 to 35\,$\mu$m. The ground-based VLT spectra spatially resolve the binary to yield separate spectra for each component. In Section 3, we discuss the results of our analysis, and model the spectra using both a simple slab model and a more detailed two-dimensional compact disk model. The analysis finds evidence for a compact source of water vapor emission in VV CrA A, with high temperature, column density and turbulent velocity. Finally, in Section \ref{sec:discussion}, we discuss possible physical origins for the water reservoir, including a circumplanetary disk, and related implications.

\begin{figure*}[ht!]
\epsscale{1.2}
\plotone{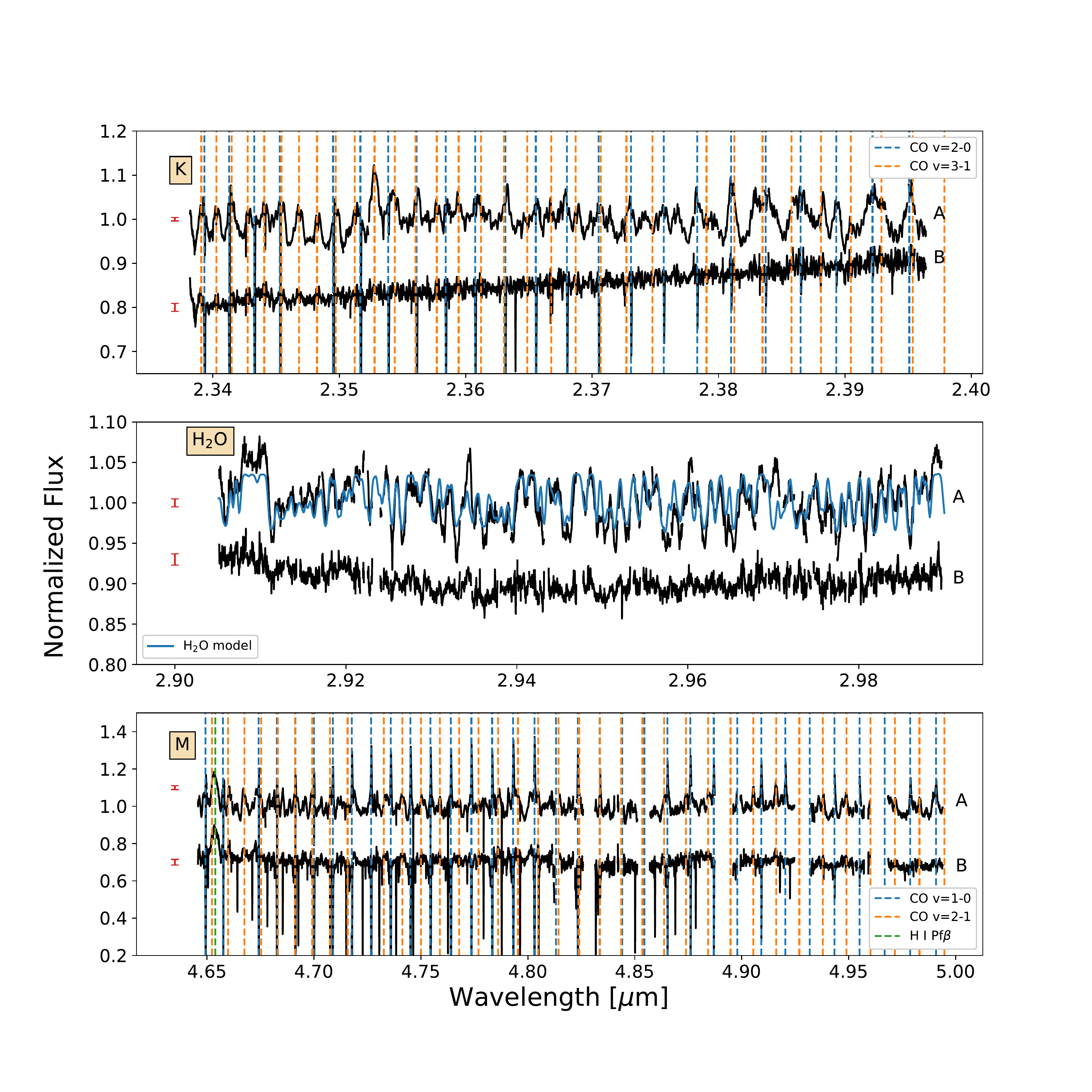}
\caption{CRIRES spectra of VV CrA A (top curves) and B (bottom curves) in the ``K'' (top), ``H$_2$O'' (middle) and ``M'' (bottom) settings.  Dashed vertical lines mark prominent $^{12}$CO and H I transitions.  Emission features from VVCrA A in the ``H$_2$O'' setting are due to numerous blended vibrational water transitions; the blue curve shows our nominal water slab model, discussed in Section \ref{sec:water_reservoir}.  Representative errors (which we estimate to be $\sim3$ times the statistical error) are shown on the left.
\label{fig:crires_overview}}
\end{figure*}

\section{Observations and Data Reduction}
\label{sec:observations}

\subsection{CRIRES 2--5\,$\mu$m high-resolution spectroscopy}
Near-infrared ($2-5\,\mu$m), high-resolution ($\lambda/\Delta\lambda\sim95,000$) spectra were obtained with the CRyogenic Infrared Echelle Spectrometer (CRIRES; \citealp{Kaeufl04}) on the European Southern Observatory Very Large Telescope (ESO VLT), as part of programs 179.C-0151 (PI:van Dishoeck) and 082.C-0432 (PI:Pontoppidan). 
%An observation log can be found in Table \ref{table:crires_log}.

\begin{deluxetable*}{lllccll}[ht!]
\tablecolumns{7}
\tablewidth{0pc}
\tablecaption{CRIRES observing log \label{table:crires_log}}
\tablehead{
\colhead{Obs. ID} & \colhead{UT Date} & \colhead{$\lambda_\mathrm{ref}$} & \colhead{PA}          & \colhead{Exp. time} & \colhead{Calibrator} & \colhead{Setting} \\
\colhead{}        & \colhead{}        & \colhead{[$\mu$m]}    & \colhead{[$^{\circ}$]}& \colhead{[s]}       & \colhead{}           & \colhead{}
}
\startdata
200165762 & 2007-04-25 & 4.7160 & 45 & 600 & BS 5646, BS 6879, BS 7236 & M\\
200165762 & 2007-04-25 & 4.7300 & 45 & 600 & BS 5646, BS 6879, BS 7236 & M\\
200165762 & 2007-04-25 & 4.8400 & 45 & 300 & BS 5646, BS 6879, BS 7236 & M\\
200168547 & 2007-08-30 & 2.9515 & 44 & 720 & BS 6084, BS 7236, BS 7337 & H$_2$O\\
200168547 & 2007-08-30 & 2.9650 & 44 & 500 & BS 7236, BS 7337          & H$_2$O\\
200168547 & 2007-08-30 & 3.7170 & 44 & 360 & BS 7236, BS 7337          & organics\\
200168597 & 2007-08-31 & 3.6100 & 44 & 600 & BS 7236                   & organics\\
200168597 & 2007-08-31 & 3.6200 & 44 & 600 & BS 7236                   & organics\\
200168597 & 2007-08-31 & 3.7040 & 44 & 600 & BS 7236                   & organics\\
200168649 & 2007-09-01 & 2.3730 & 44 & 800 & BS 7236, HIP 101589       & K\\
200168649 & 2007-09-01 & 2.3780 & 44 & 560 & BS 7236, HIP 101589       & K\\
200168649 & 2007-09-01 & 4.7700 & 44 & 960 & BS 7236                   & M\\
200168649 & 2007-09-01 & 4.7795 & 44 & 960 & BS 7236                   & M\\
200169023 & 2007-09-06 & 3.4565 & 44 & 360 & BS 7913                   & organics\\
200169023 & 2007-09-06 & 3.5350 & 44 & 360 & BS 7913                   & organics\\
200179090 & 2008-08-10 & 4.9462 & 44 & 480 & BS 7236                   & M\\
\enddata
\end{deluxetable*}

Data were reduced using IDL routines described in \citet{Pontoppidan11}. Reduction steps included nod subtraction, trace linearization, optimal extraction, and wavelength calibration with telluric emission lines.   All wavelengths were converted to heliocentric values. Telluric calibration was performed with division by observed A- and B-type star telluric calibrators, whose near-IR spectra are relatively featureless. As a final step, spectra in adjacent wavelength settings were combined together into one of four settings labeled: ``K'', ``H$_2$O'', ``organics'', or ``M''.  The complete wavelength coverage of the four settings is 2.3382--2.3964$\,\mu$m, 2.9051--2.9897$\,\mu$m, 3.3983--3.7495$\,\mu$m and 4.6457--4.9940$\,\mu$m, in that order. The ``H$_2$O'' and ``organics'' settings are both in the ``L'' filter, so we refer to these collectively as ``L-band'' spectra in the remainder of the text.  An observing log for the CRIRES spectra is provided in Table \ref{table:crires_log}. 

Due to the narrow slit (0\farcs2) of CRIRES, the spectra cannot be accurately calibrated to an absolute flux density scale. Consequently, we scale the VV CrA A spectra to WISE photometry (24 W1 measurements of the sum of the two VV CrA components between April and September 2010, average W1=4.38\,mag), coupled with a flux density ratio between VV CrA A and B at $L'$ of 1.01 \citep{Sullivan19}, obtained on July 12, 2015. However, it is known that VV CrA is variable in the mid-infrared at the $\sim30\%$ level on a 10-year time scale \citep{Kospal12}. The WISE light curve indicates variability of 20\% on a 6-month time scale.  

A representative overview of the CRIRES spectra is shown in Figure \ref{fig:crires_overview}, and the full data set is displayed in an appendix (Figures \ref{fig:crires_A_K}--\ref{fig:crires_A_M}).\footnote{Reduced data are also available at https://www.stsci.edu/$\sim$pontoppi/\#data.}  M-band data for VV CrA B were previously analyzed by \citet{Smith09} and M-band data for both sources were previously discussed by \citet{Bast11}. 

\subsection{VISIR 12\,$\mu$m high-resolution spectroscopy}
VV CrA was observed with the VISIR spectrograph \citep{Lagage04,Kaufl15} on the VLT, as part of the large program 198.C-0104 (PI:Pontoppidan). Spectra were obtained using the high-resolution mode, with a resolving power of $R\sim 30,000$.

\begin{figure*}[ht!]
\epsscale{1}
\plotone{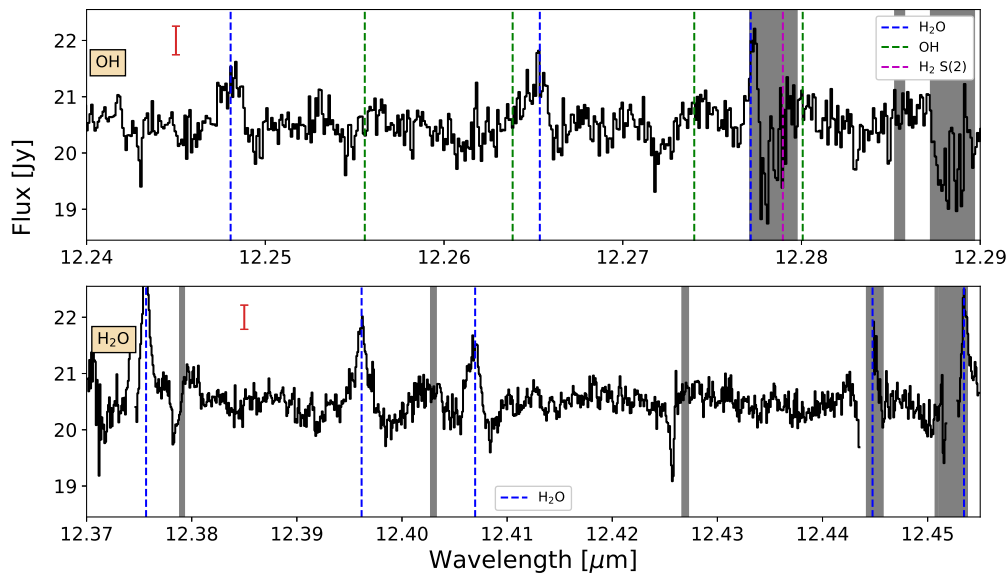}
\caption{Reduced spectra of VV CrA A in the two water/OH settings observed with VISIR. Gray rectangles mark regions with atmospheric transmission less than 80\%.  Dashed lines mark the theoretical locations of H$_2$O, OH and H$_2$ transitions.  Data wavelengths are heliocentric and theoretical wavelengths have been shifted by $-$1 km s$^{-1}$, the VV CrA A stellar velocity \citep{Fang18}.  In the upper left, we show representative error bars computed from the standard deviation of flat spectral regions. \label{fig:VISIR_overview}}
\end{figure*}

Both components of VV CrA were observed in April 2017 in three wavelengths settings: one setting centered at 12.414\,$\mu$m, covering rotational water lines, one centered at 12.27\,$\mu$m containing OH and the H$_2$ S(2) line, and one centered on 12.837\,$\mu$m containing the [Ne II] line.  The [Ne II] data were previously analyzed in \citet{Pascucci20} and will not be discussed in this work. Asteroids (6) Hebe, and (10) Hygiea were observed as telluric calibrators. For the April 2017 observations of VV CrA, the slit was oriented along the binary position angle (PA) and both targets were observed simultaneously. On 2017 September 08, VV CrA was observed again at two different position angles, separated by 180 degrees, in order to perform spectro-astrometry. An observing log for VISIR is shown in Table \ref{table:visir_log}.

\begin{deluxetable*}{lllccll}[ht!]
\tablecolumns{7}
\tablewidth{0pc}
\tablecaption{VISIR observing log \label{table:visir_log}}
\tablehead{
\colhead{Obs. ID} & \colhead{UT Date} & \colhead{$\lambda_0$} & \colhead{PA}          & \colhead{Exp. time} & \colhead{Calibrator} & \colhead{Setting} \\
\colhead{}        & \colhead{}        & \colhead{[$\mu$m]}    & \colhead{[$^{\circ}$]}& \colhead{[s]}       & \colhead{}           & \colhead{}
}
\startdata
1732798 & 2017-04-25 & 12.414 & 136 & 1890 & (6) Hebe    & H$_2$O \\
1733361 & 2017-04-26 & 12.270 & 136 & 1800 & (6) Hebe    & OH     \\
1733364 & 2017-04-26 & 12.837 & 0   & 1800 & (6) Hebe    & [Ne II]
\tablenotemark{a}\\
1830517 & 2017-09-08 & 12.414 & 136 & 2160 & (10) Hygiea & H$_2$O \\
1830520 & 2017-09-08 & 12.414 & -44 & 2160 & (10) Hygiea & H$_2$O \\
\enddata
\tablenotetext{a}{See \citet{Pascucci20}.}
\end{deluxetable*}

Spectra were reduced using Python code developed by our team. Reduction steps included nod subtraction, trace linearization, optimal extraction, and wavelength calibration using telluric emission lines. To remove telluric absorption features, both source and standard spectra were fit by a {\tt molecfit} \citep{Smette15,Kausch15} model, which was subsequently divided into each spectrum. The atmospheric-corrected source was then divided by the atmospheric-corrected standard to produce the final spectrum. The use of an atmospheric model helps account for differences in telluric absorption between source and standard, while the use of the asteroid as telluric calibrator corrects for any instrumental effects, as described further in \citet{Mandell11}.  Finally, spectra were fringe-corrected using low-order polynomials, and flux calibrated using the flux density measured through the Si-6 (12.33 $\mu$m) VISIR bandpass from \citet{Kruger11}. Figure \ref{fig:VISIR_overview} shows the reduced VISIR spectra in the three settings.  

Spectro-astrometric (SA) measurements in the 12.414 $\mu$m  water setting are shown in Figure \ref{fig:h2o_sa}. A detailed discussion of the SA method can be found in \citet{Pontoppidan11}. In short, spectro-astrometry is a measurement of the spatial centroid of the emission as a function of wavelength.  This spatial centroid is expected to vary with wavelength, as the blueshifted part of a Keplerian disk is spatially offset from the redshifted part.  Since the measured location of the centroid is influenced by instrument- and sky-induced residuals, the object is observed at two anti-parallel position angles, and the calibrated spectro-astrometric signal is constructed by subtracting the SA offsets of the two position angles (in this case, 136$^\circ$ and $-44^\circ$) and dividing by two. With this method, instrument- and sky-induced residuals cancel out, and only source-based signals remain. Based on empirical tests aimed at maximizing the signal-to-noise ratio, we used a 12 pixel-wide box to calculate the spatial centroid. Because the SA extraction box truncates the PSF, an empirically-determined correction factor of 1.2 was applied to the final SA spectrum.  

\subsection{Spitzer 10--35\,$\mu$m medium-resolution spectroscopy}
\begin{figure*}[ht!]
\epsscale{1}
\plotone{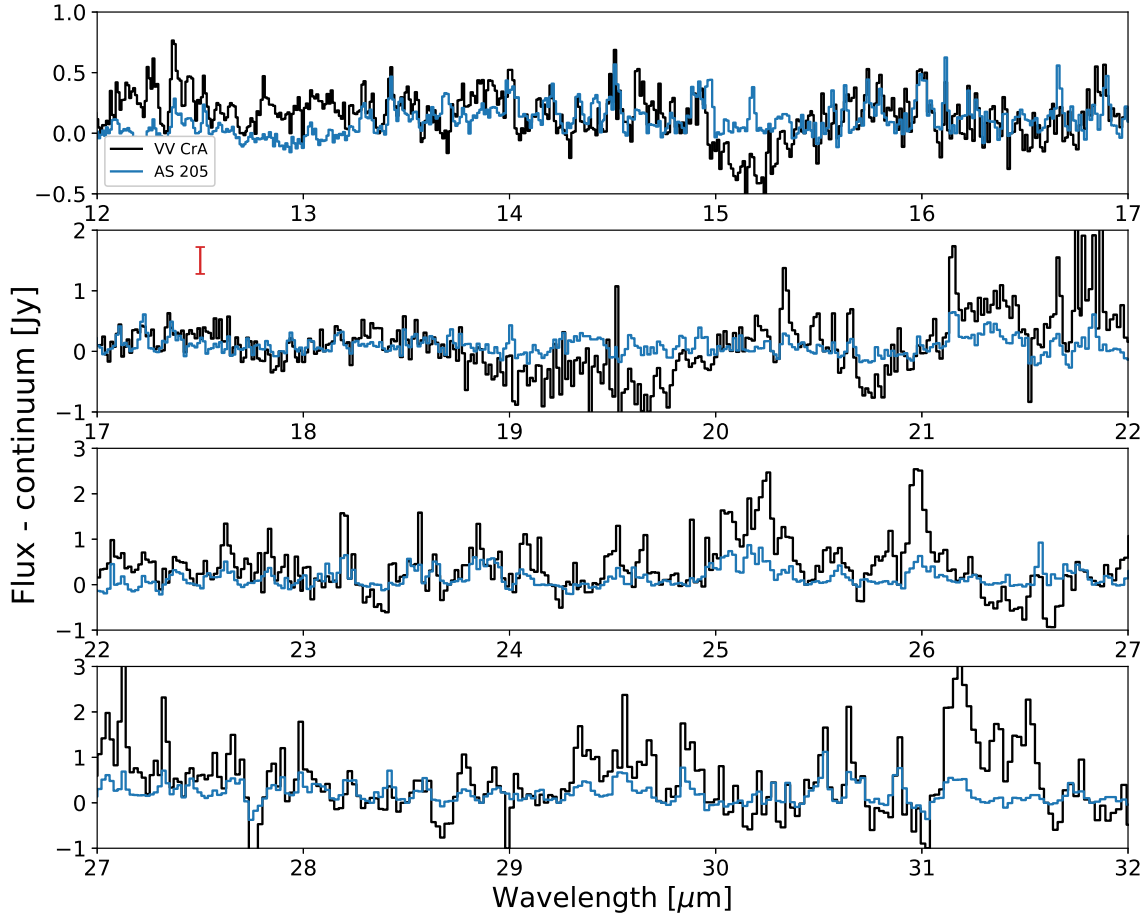}
\caption{Continuum-subtracted Spitzer-IRS high-resolution spectra of VV CrA (A \& B combined; black) and the water-rich disk AS 205 (blue; \citealp{Pontoppidan10a}), for comparison.  A representative error bar is shown in the second panel, computed from the standard deviation between 17 and 18 $\mu$m. \label{fig:spitzer}}
\end{figure*}

VV CrA was observed with the Spitzer InfraRed Spectrometer \citep[IRS, ][]{Houck04} in the 9.9--19.6 $\mu$m short-high (SH) and 18.7--37.2 $\mu$m long-high (LH) modules as part of program 20611 (PI: Wright). In this work, we use the archival data from program 20611 (AOR 14921984), but re-reduced following the steps described in \citet{Pontoppidan10a}. The VV CrA binary was not spatially resolved by IRS, so the spectra represent the sum of the two targets. The continuum-subtracted IRS SH and LH spectra are shown in Figure \ref{fig:spitzer}.

\section{Analysis}
\label{sec:analysis}
\subsection{Evidence for an unusual compact, hot and dense water reservoir}
\label{sec:water_reservoir}
The CRIRES $L$-band spectrum of VV CrA A (Figures \ref{fig:crires_overview},\ref{fig:crires_A_H2O} and \ref{fig:crires_A_org}) is characterized by a dense forest of lines from ro-vibrational transitions of water vapor. While the observed spectrum covers only parts of the atmospheric $L$-band window, water lines are observed at nearly all covered wavelengths, including the range around 3\,$\mu$m, where water vapor emission is commonly seen in protoplanetary disks \citep{Banzatti17}, but also out to the red end of the spectral coverage at 3.75\,$\mu$m. The extension of the 3$\,\mu$m water stretching-mode band to such long wavelengths is unusual, appearing similar to only one previously studied disk, V1331 Cyg \citep{Doppmann11}, which we discuss further in Section \ref{sec:discussion}.  Overall, we inspected 37 $L$-band spectra of protoplanetary disks and did not note another similar spectrum. \footnote{The spectra had a range of SNR's and a range of line-to-continuum ratios for water emission, including non-detections, so the detection rate should not be taken to be the same as the occurrence rate.}

The spectral extent of the observed water band in VV CrA A suggests that unusually high temperatures and/or unusually high column densities are present. Further, the fact that the water lines have roughly constant strength ($\sim$0.1--0.2\,Jy) across a band that includes transitions with a wide range of upper level energies ($\sim$5,000--10,000\,K) and strengths ($<$0.01--90\,s$^{-1}$) suggests that they are optically thick, which would also imply high column densities.

In the following, we investigate the properties of the emitting material required to reproduce the observed water spectrum in VV CrA A, and discuss which unusual circumstances may be required to explain it. We use both a simple slab model as well as a more detailed two-dimensional radiative transfer model to estimate gas temperature, columns density/abundance, and emitting area. 

\subsection{Slab model of the VV CrA A water emission}
\label{sec:slab_models}
We investigated the VV CrA A ``H$_2$O'' spectrum with the use of a simple slab water emission model (``slabspec''; \citealp{Salyk20}), which consists of a slab of water vapor with a single temperature ($T$), column density ($N_\mathrm{H2O}$) and emitting area ($A$). Emission line data come from the HITRAN database \citep{Gordon17}. The model also has a tunable parameter setting the standard deviation of the local Gaussian line broadening ($v$), which serves as a parameterization of the local turbulent velocity. We then convolved all of the model spectra with a Gaussian with a Full Width at Half Maximum (FWHM) of $\sim$30 km s$^{-1}$ ($\sigma=12.7$ km s$^{-1}$) to match the appearance of the observed spectra.   This step represents non-local broadening due to Keplerian motion and, potentially, non-Keplerian outflows or winds.  (A more detailed investigation of line shapes is described later in Section \ref{sec:lineshape_models}).

\begin{figure*}
\epsscale{1.1}
\includegraphics[width=1\textwidth]{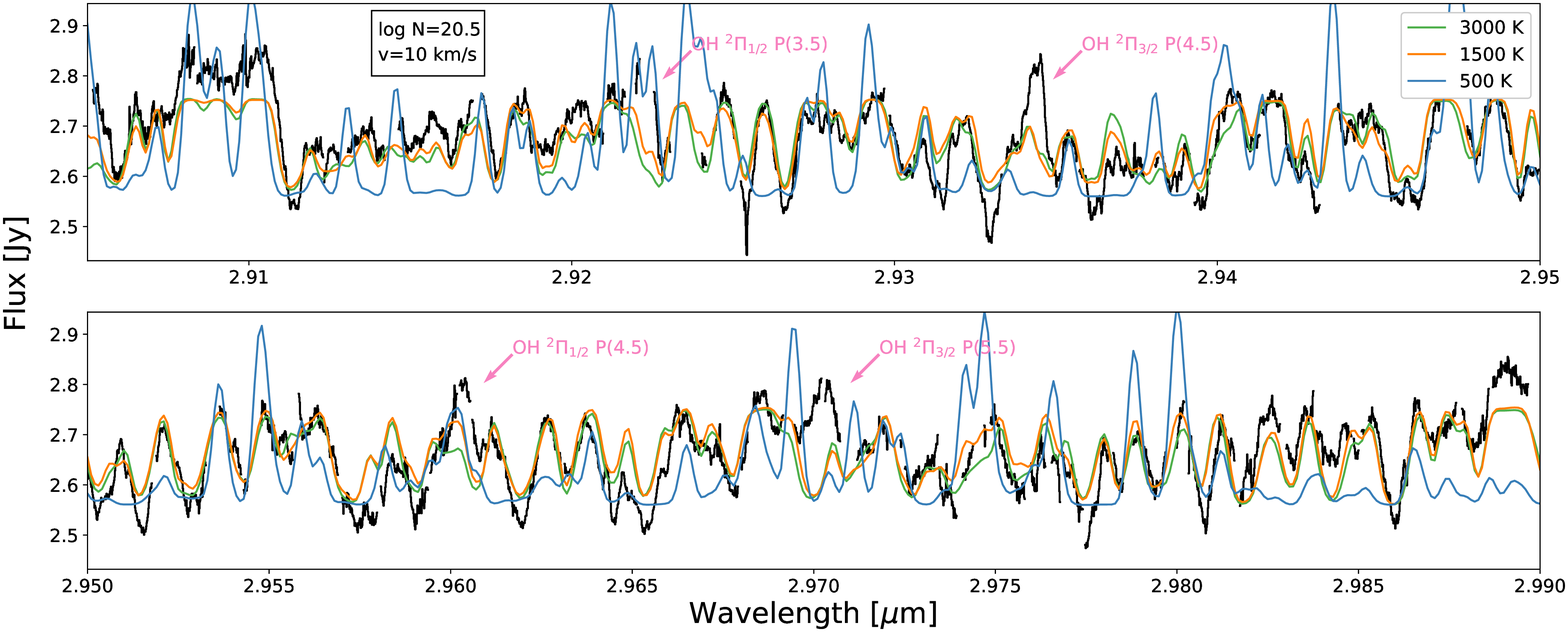}
\includegraphics[width=1\textwidth]{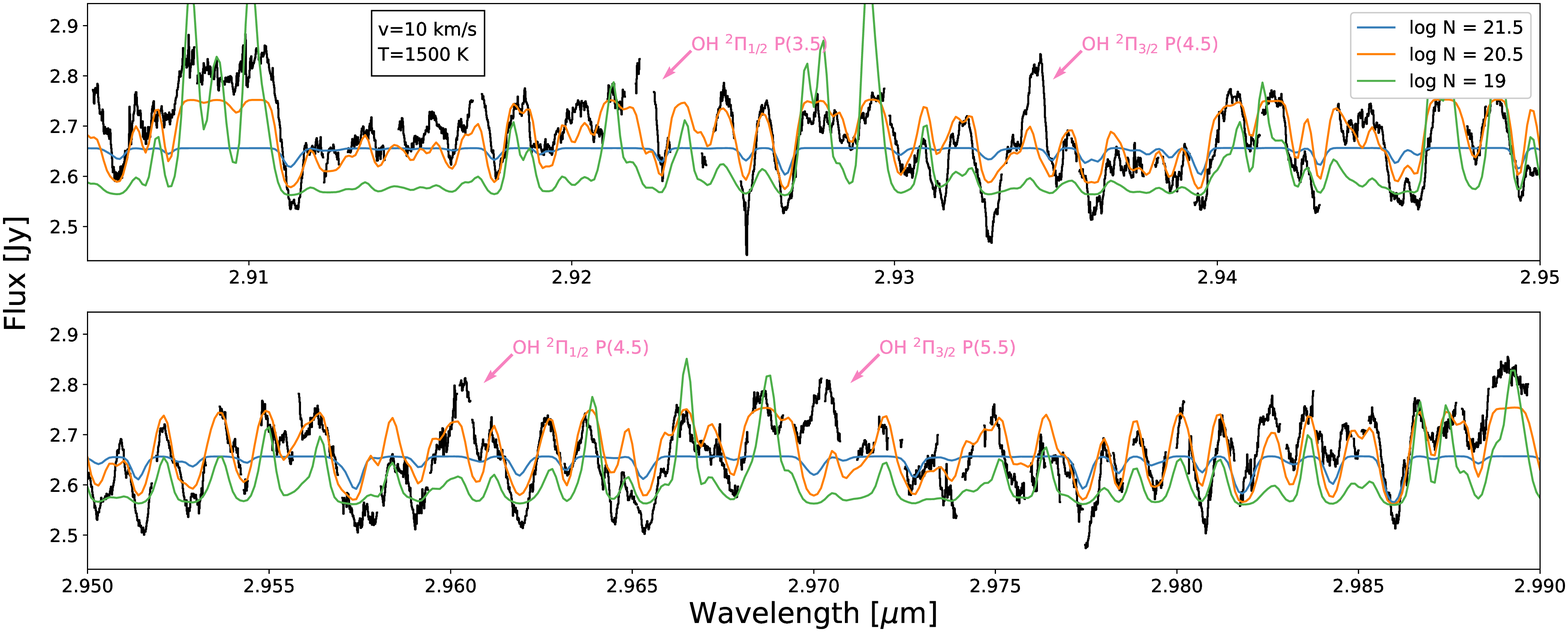}
\includegraphics[width=1\textwidth]{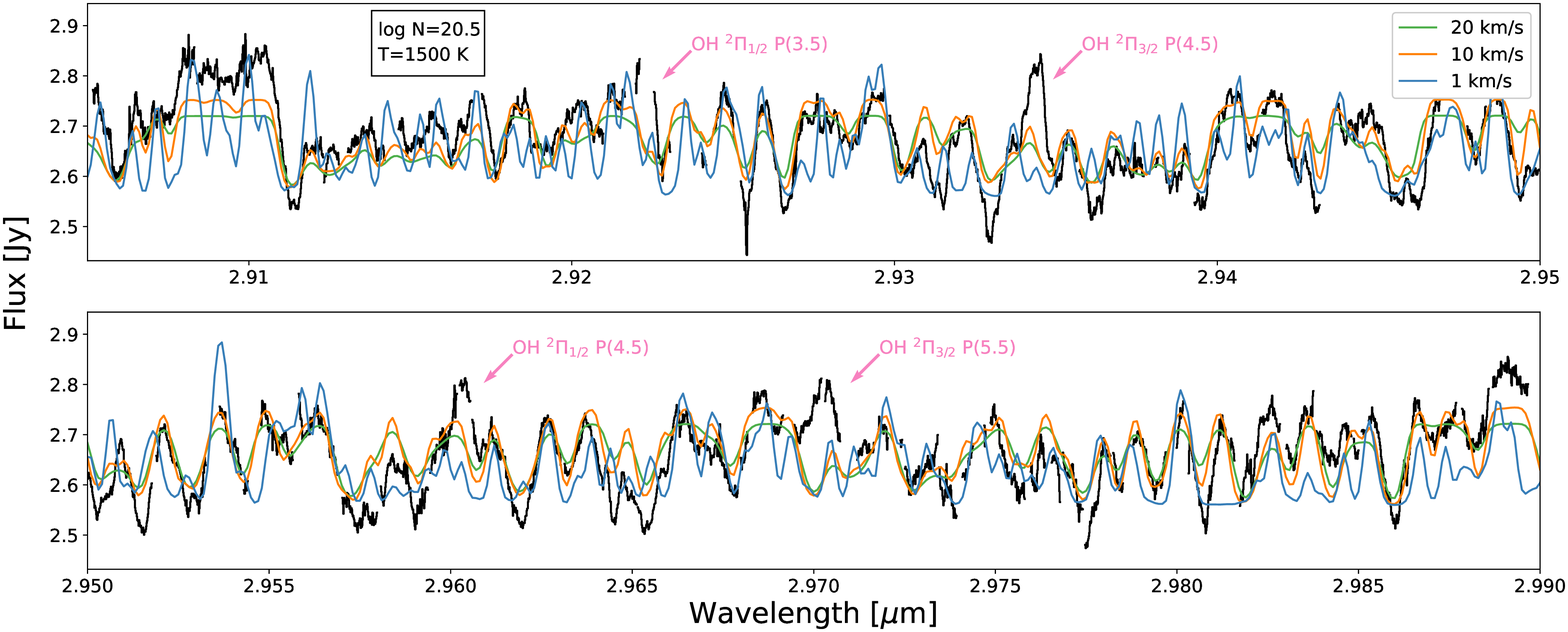}
\caption{Water vapor slab models with different values of temperature (top), water column density (middle), and line broadening (bottom), as compared to CRIRES data for VV CrA A in the ``H$_2$O'' setting.  Arrows mark probable OH emission features.  Boxes in the upper left list fixed model parameters.
\label{fig:CRIRES_h2o_models}}
\end{figure*}

\begin{figure*}
\epsscale{1.1}
\includegraphics[width=1\textwidth]{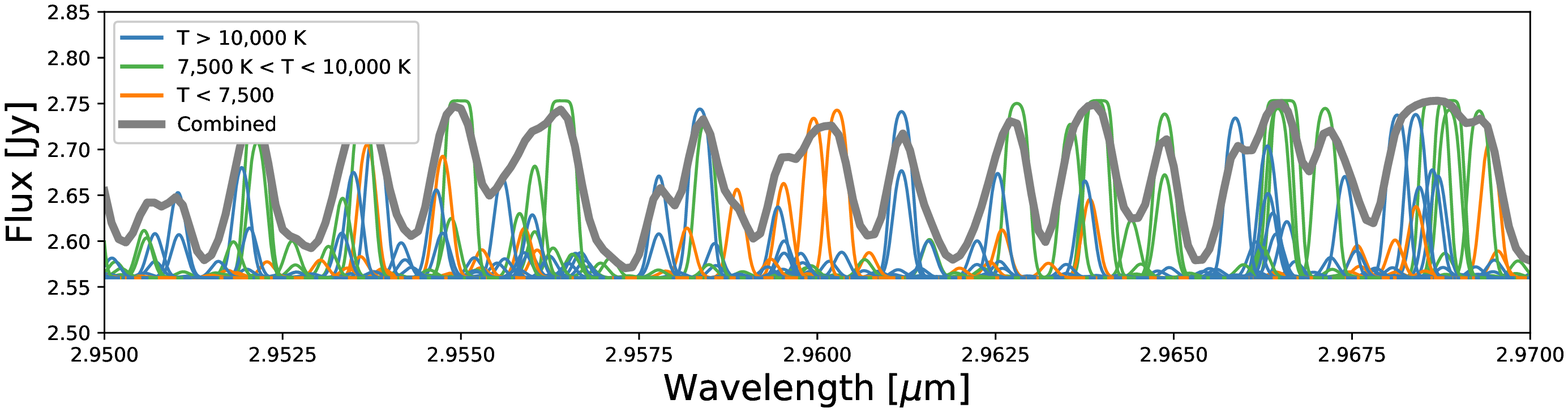}
\caption{Breakdown of a portion of our nominal model into its component individual emission lines. Color coding shows the excitation temperature of the upper levels.  The thick gray curve shows the combined, convolved model.  \label{fig:model_breakdown}}
\end{figure*}

We explored a range of values for $N_\mathrm{H2O}$, $T$ and $v$ to derive a nominal model that fits the ro-vibrational water lines in the ``H$_2$O'' setting. Since changes to the emitting area scale the entire spectrum linearly by a constant factor, we adjusted this factor after choosing $N_\mathrm{H2O}$, $T$, and $v$, such that the value of emitting area minimized the sum of squared residuals.  Our nominal model is taken to have $T=1500\ $K, $N_\mathrm{H2O}=3\times10^{20}\ \mathrm{cm}^{-2}$, $v=10\ \mathrm{km\ s}^{-1}$ and $A = 0.003$ AU$^2$ --- see Figure \ref{fig:CRIRES_h2o_models}, orange curves.  Throughout the remainder of the text, we refer to this as our nominal L-band slab model.  Figure \ref{fig:model_breakdown} shows a breakdown of the nominal model into contributing emission lines over a portion of the spectrum, demonstrating the high excitation temperatures of the emitting lines, and the complexity of the resulting spectra.

Figure \ref{fig:CRIRES_h2o_models} shows how each of the free parameters affects the model. Temperature and column density both affect the relative line strengths, with higher temperatures adding flux in higher-excitation lines and higher column densities adding flux in weaker lines. Column density and local line broadening both affect the broadness of optically thick lines; an increased column density increases optical depths at all velocities, while an increased local line broadening redistributes molecules to a wider range of velocities. However, the effects of increasing column density and local line broadening are different --- at high column densities ($N_\mathrm{H2O}\sim10^{23}\ \mathrm{cm}^{-2}$), models approach the flat blackbody optically thick limit for all wavelengths, while with large line broadening, optically thick lines develop flat shapes, while optically thin lines only broaden slightly. 

\begin{figure*}
\epsscale{1}
\plotone{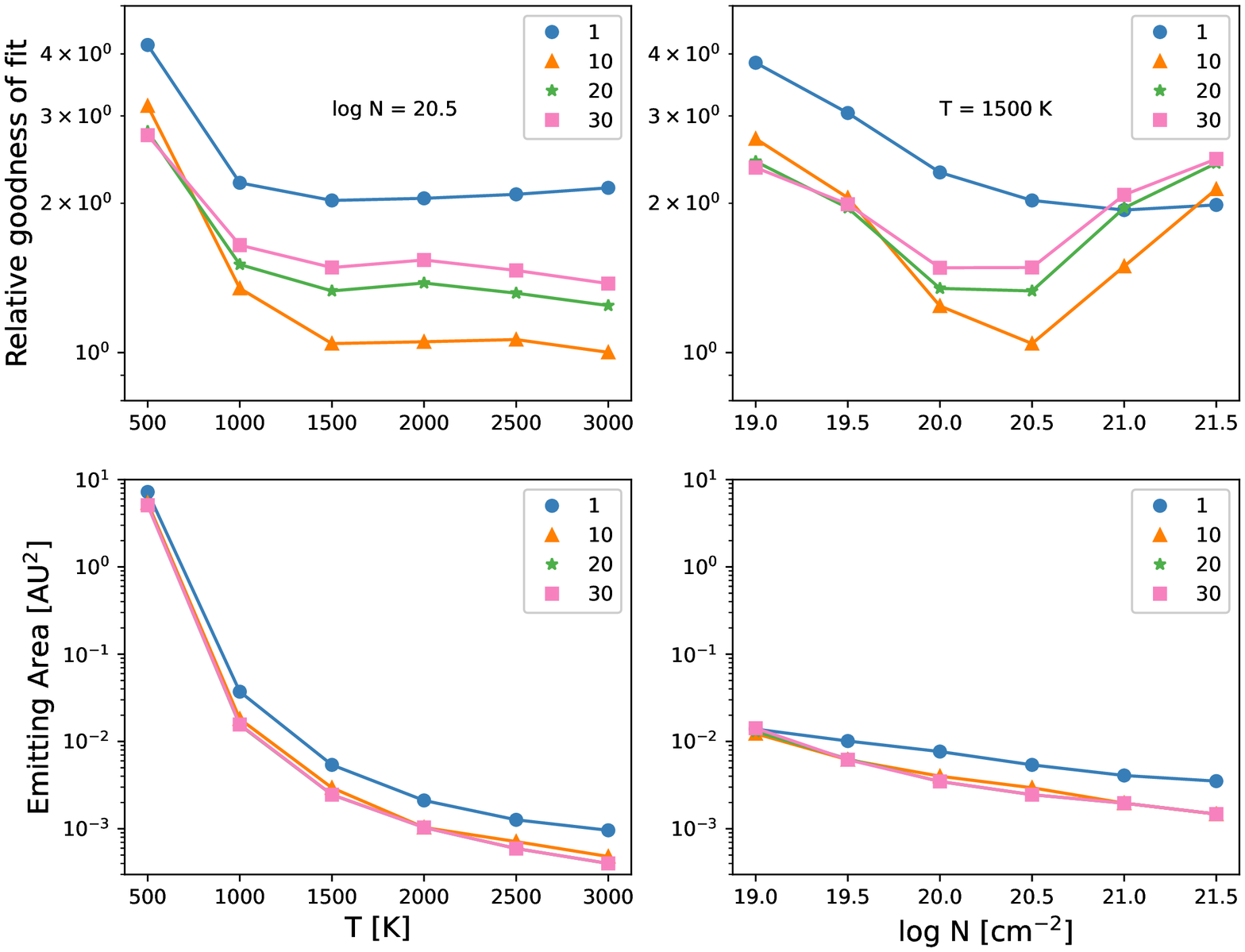}
\caption{Top: Relative goodness-of-fit (residual sum of squares divided by that of the best-fit model) for a range of model parameters.  The top left plot shows the dependence of goodness-of-fit on temperature, with $N_\mathrm{H2O}$ fixed at $3\times10^{20}\ \mathrm{cm}^{-2}$, while the top right plot shows the dependence on column density, with $T$ fixed at $1500\ $K.  Different colors and symbols show different values of the local line broadening, $v$, in km s$^{-1}$.  Bottom: Corresponding emitting areas for each model.
\label{fig:chisq_plot}}
\end{figure*}

Figure \ref{fig:chisq_plot} shows the relative goodness of fit for a model grid with different values of $N_\mathrm{H2O}$, $T$ and $v$, where we define the relative goodness of fit as the sum of squared residuals divided by that of the best-fit model.  (None of the models can perfectly match all of the spectral features, so reduced-$\chi^2$ values remain high for all models.)  Residuals were computed between 2.905 and 2.99 $\mu$m, excluding regions near known OH emission lines, and for each set of $N_\mathrm{H2O}$, $T$ and $v$, we adjusted the emitting area to minimize the sum of squared residuals.   We see that high temperatures ($T\gtrsim$1500 K) and high local line broadening ($v\gtrsim$10 km s$^{-1}$) are preferred, but the upper limits of these parameters are not well constrained.  Best-fit models have column densities of $N_\mathrm{H2O}=1-3
\times10^{20}$ cm$^{-2}$.  Associated emitting areas are $\lesssim 0.005$ AU$^2$, with smaller emitting areas corresponding with higher temperatures, and vice versa.

Note that the best-fit emitting area is not necessarily equal to the size of the region containing water vapor, as lower temperature water vapor will not effectively emit at these wavelengths. Furthermore, when our nominal $L$-band slab model is applied to the broader $L$-band wavelength range (i.e., the wider ``organics'' setting), it predicts emission weaker than that observed.  When extrapolated to the longer wavelengths covered by VISIR and Spitzer, it also greatly underpredicts observed emission lines. This is further evidence that water vapor exists at a range of temperatures.  We consider this further in Sections \ref{sec:radlite_model} and \ref{sec:spitzer_visir}.

\subsection{Velocity of the water vapor}
\label{sec:lineshape_models}
The velocities of the $L$-band emission lines can be used to constrain the emitting location of the hot, compact emitting region. Hot emission lines can naturally be produced near the inner edge of a protoplanetary disk. However, as discussed in Section \ref{sec:slab_models}, convolution with a FWHM$\sim$30 km s$^{-1}$ ($\sigma=12.7$ km s$^{-1}$) line shape  matches the appearance of the data.  As we will discuss, these widths may be too narrow to explain with a standard protoplanetary disk model.

\begin{figure*}
\epsscale{1.2}
 \plotone{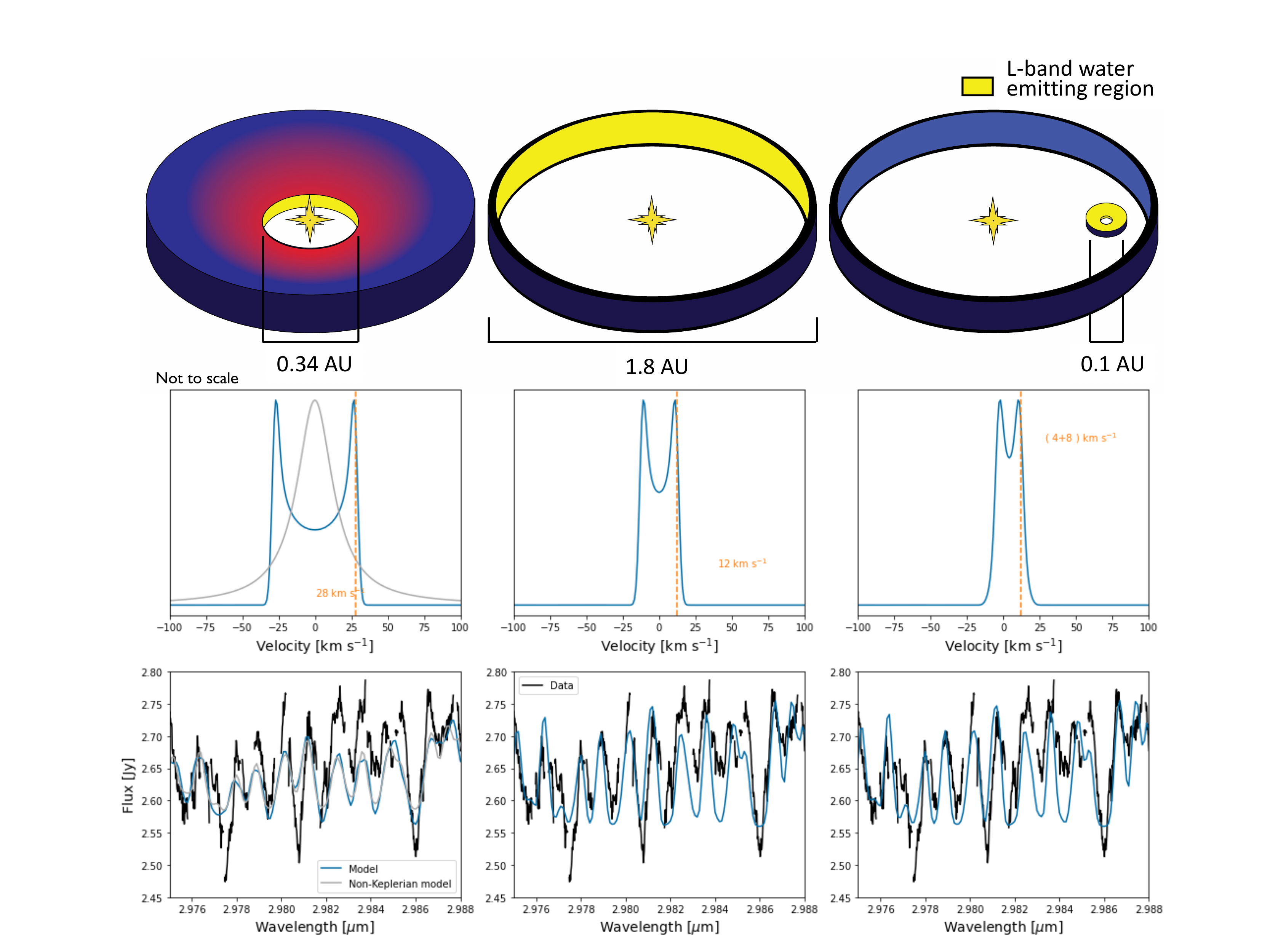}
 \caption{Representative drawings (top - not to scale), line shapes (middle), and resultant spectra (bottom) for the three L-band water emission models discussed in Section \ref{sec:lineshape_models}.  Left: A ring at the dust inner radius of 0.17 AU, surrounding a 0.55 M$_\odot$ star with a Keplerian (blue) or non-Keplerian Lorentzian (gray) line profile.  Middle: A ring at 0.9 AU, surrounding a 0.55 M$_\odot$ star.  Right: A circumplanetary disk surrounding a 6 M$_J$ planet, offset by 4 km s$^{-1}$.  All models are inclined by 32$^\circ$.  \label{fig:lineshape_cartoon}}
\end{figure*}

In Figure \ref{fig:lineshape_cartoon}, we show how the location and shape of the compact emitting region affect the appearance of the observed spectrum. We constructed a model from a ring of constant irradiance (i.e., equivalent to a single temperature blackbody) with emitting area $0.003$ AU$^2$, which we then oriented at an inclination of 32$^\circ$ \citep{Gravity21} to produce a line shape. To compare with the data, we then took the (unconvolved) spectrum output by our nominal slab model (see Section \ref{sec:slab_models}) and convolved it with the line shape model.

In the first model, as shown in the top left panel of Figure \ref{fig:lineshape_cartoon}, we considered a ring beginning at the measured dust inner radius of $\sim$0.17 AU \citep{Gravity21}, extending to 0.173 AU (to match the slab-model emitting area), orbiting a 0.55 M$_\odot$ star.  This model produces a wide, double-peaked line profile; when convolved with the slab model, it flattens out the spectrum such that it does not well match the data.  It has been observed by e.g. \citet[][]{Bast11} that molecular disk emission line profiles are sometimes non-Keplerian, so we also consider a Lorentzian profile matching the observed CO M-band line shape, with line wings consistent with inner disk velocities. This, too, produces a spectrum flatter than observed.

In a second model (Figure \ref{fig:lineshape_cartoon}, middle panel), we arbitrarily set the disk inner radius to 0.9 AU to match the observed line widths. The narrower resulting line profile matches the observed data when convolved with the slab model; however, it requires the adoption of a very thin ($\Delta r=6\times10^{-4}$\,au), hot ring to match the slab model emitting area. The ring location conflicts with the measured dust rim location of 0.17$\pm$0.02 AU \citep{Gravity21}.  But perhaps the high UV radiation flux around this rapidly accreting ($\dot{M}=3\times10^{-7} M_\odot\,\mathrm{yr}^{-1}$; \citealp{Fang18}) star causes photodissociation of H$_2$O out to larger radii \citep{Bethell09}.  It is unclear, however, how to produce such a narrow ring of hot water, and it may be difficult to reconcile the high gas temperatures with the expected equilibrium temperature of 470 K at 0.9 AU around a 4.9 $L_\odot$ star.

As a third model, we considered the idea that the narrow lines, high column density, and small emitting area might be explained by a {\it circumplanetary} disk (CPD) around a massive, young planet or substellar companion. To test this option, we constructed a model for a disk around a 6 M$_J$ planet, with an inner radius of 10 R$_J$ extending to 60 R$_J$ (discussed further in Section \ref{sec:radlite_model}). As shown in Figure \ref{fig:lineshape_cartoon} (bottom panel), a CPD also produces a narrow line profile consistent with the data, and will match the emitting area with a more plausible range of disk radii. This concept is developed further with a two-dimensional model in Section \ref{sec:radlite_model}.

One important, potential difference between a CPD and a narrow ring within a protoplanetary disk is that the CPD lines will be offset due to the Keplerian velocity of the planet. The magnitude of the offset may range from 0 (if the planet is on the far or near tangent of its orbit), up to $v_{\rm CPD, Kepler} * \sin(i)$, where $i$ is the inclination of the CPD orbit around VV CrA A. As a representative example, we show an offset of $\sim 4\,\rm km\,s^{-1}$, given by the Keplerian speed around a 0.55 M$_\odot$ star of a ``Jovian'' orbit at 5 AU, projected to $i=32^{\circ}$ and an azimuth of 45$^{\circ}$. The actual semi-major axis of any putative CPD is unknown.

To test this idea, we compared the L-band water emission to the stellar velocity.  \citet{Fang18} find a heliocentric stellar velocity of $-1\pm0.3$ km s$^{-1}$ for VV CrA A.  For the $L$-band data, lines cannot be separated, so we estimated the velocity by two methods: minimization of $\chi^2$ between our slab model and data, which finds a best-fit velocity of 4 km s$^{-1}$ and a cross-correlation between slab model and data, which suggests a velocity of $-1$ km s$^{-1}$.  Therefore, given systematic errors of a few km s$^{-1}$ we find no significant difference between the stellar, and L-band water, velocities.  Future work would need to more carefully reduce systematics to check for velocity offsets consistent with a CPD.  We note, however, that offsets greater than 5--10 km s$^{-1}$ with respect to the stellar velocity are not consistent with our data.
 
In summary, we show that the observed fluxes and velocities are consistent with either a thin ring far from the star, or with a compact spot, e.g. a circumplanetary disk.  We consider the CPD hypothesis further in the next section.

\subsection{Two-dimensional radiative transfer model of a circumplanetary disk}
\label{sec:radlite_model}

In this section, we discuss more sophisticated modeling we performed to further explore the possibility that the $L$-band emission arises in a CPD.  We utilized the radiative transfer modeling code RADMC \citep{Dullemond04} coupled with the line radiative transfer code RADLite \citep{Pontoppidan09}. As indicated by the slab model, the water emission appears to be formed in a small emitting area with high excitation temperature, large column densities, and a high degree of turbulent broadening.  

\begin{figure*}
 \epsscale{1.3}
 \plotone{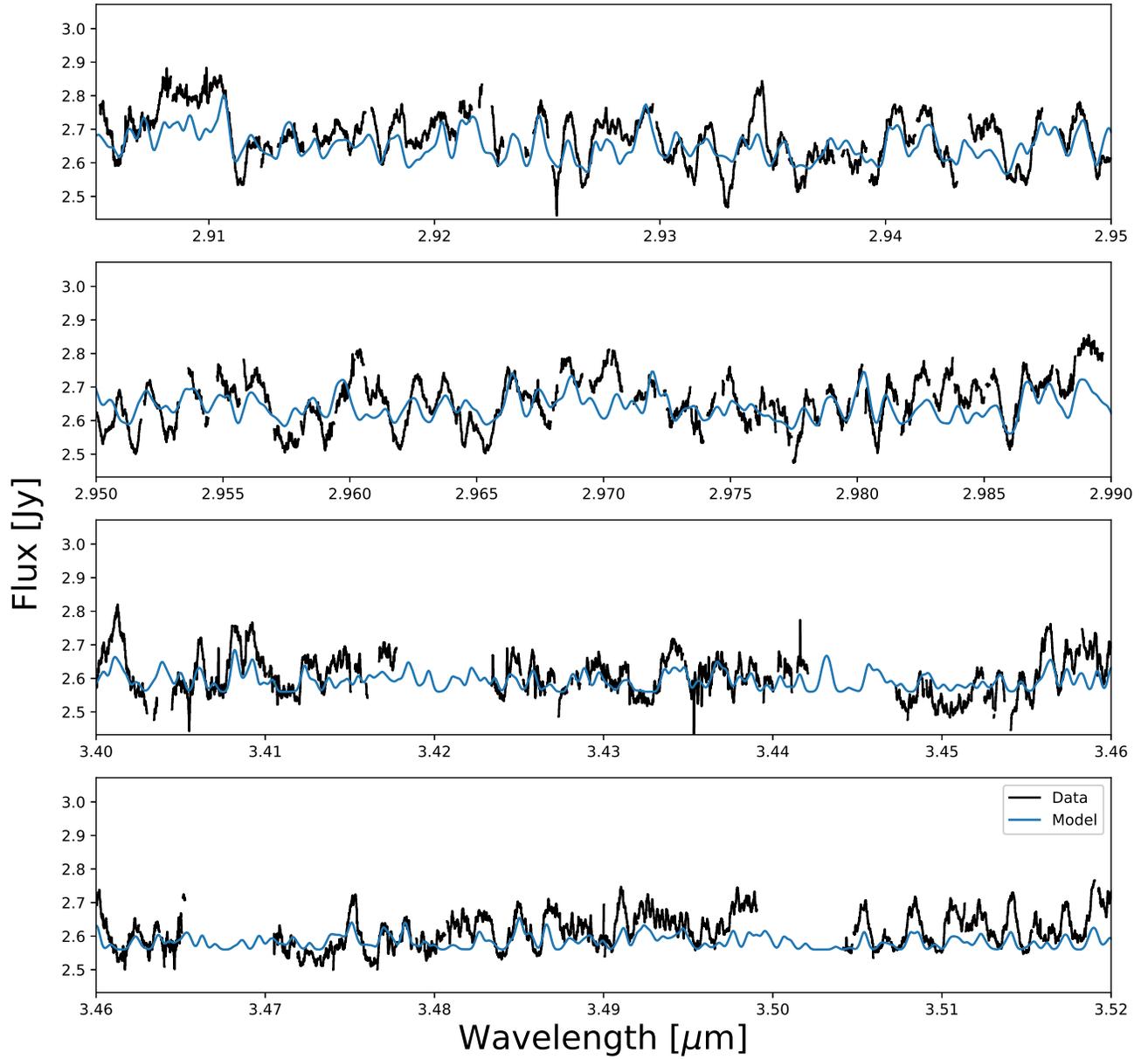}
 \caption{Comparison between a portion of the L-band spectrum of VV CrA A and a RADLite model of a 0.2 AU-sized disk around a substellar object of mass 6 M$_J$ and luminosity 0.023\,$L_{\odot}$ (see Section \ref{sec:radlite_model}). \label{fig:radlite_cpd}}
\end{figure*}

Figure \ref{fig:radlite_cpd} shows a portion of a nominal RADLite model that provides a good match to the $L$-band data. The full wavelength coverage can be seen in Figures \ref{fig:crires_A_K}--\ref{fig:crires_A_M}. The RADLite model calculations of the full spectral range are too time-consuming to investigate uniqueness, so this represents one plausible solution. 

In this model, the disk surrounds a 6 M$_J$ planetary companion. The radius and effective temperature of the central object determine its luminosity, which affects the strength of the emission lines, but does not otherwise change the appearance of the spectrum. We find that a CPD surrounding a 5 R$_J$ planet, with an effective temperature of 3200 K, corresponding to a luminosity of 0.023\,$L_{\odot}$, can generate an H$_2$O spectrum with similar line strengths to the observed spectrum.  However, other combinations of radius and effective temperatures are similarly allowed as long as the total luminosity is kept roughly constant. The disk is required to have a very high gas-to-dust ratio of 70,000 and high water abundance relative to H$_2$ of 0.05 to reach the observed optical depths. The disk is assumed to have an inner radius of 10 R$_J$ (equivalent to an effective temperature of 1600 K), and a total mass of 6$\times10^{-3}$M$_J$ or 1.9 M$_\oplus$.  We also utilized a turbulent broadening equal to 5 times the sound speed.  Following the prescription in \citet{Dullemond04}, the surface density is given by a split power law with exponent of -1 out to 0.2 AU and -12 from 0.2 to 0.5 AU. The steep index in the outer portion effectively provides a ``smooth'' disk edge at 0.2 AU. We chose a scale height ($H/R=0.2$) at 0.2\,au, with H/R scaling as $R^{1/7}$ for a slightly flared disk, and an opacity given by a mix of amorphous and crystalline silicates plus organics. 

The model assumes that the gas temperature is coupled to that of the dust, which may not be a good approximation \citep{Kamp04}. It is possible that the gas is heated above that of the dust through non-thermal heating processes both due to a strong UV field from accretion onto the planet, the central star, or a combination, thereof.  Higher gas temperatures would result in stronger lines. At the same time, densities may be too low to excite the lines through collisions with atomic and molecular hydrogen, leading to sub-thermal line populations and weaker overall line intensities \citep{Meijerink09}. In the RADLite model, such effects are neglected and local thermodynamic equilibrium is assumed. Effectively, the assumptions of coupled gas and dust temperatures with LTE excitation tend to cancel each other out and may produce similar line fluxes to the combined assumptions of decoupled gas temperatures plus non-LTE excitation \citep{Meijerink09, Pontoppidan22}. Therefore, the inferred luminosity of the putative planet/substellar companion is plausible, but is likely associated with a large uncertainty.  Similarly, model column densities do not necessarily reflect true column densities.

\begin{figure*}
 \epsscale{1.3}
 \plotone{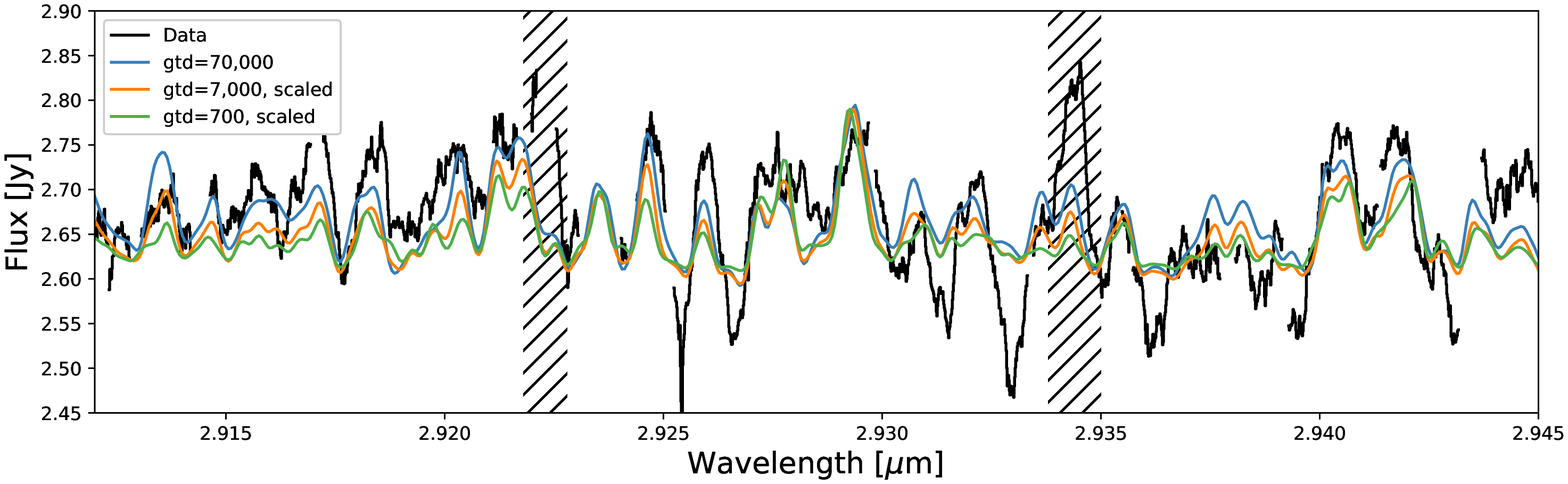}
 \caption{RADLite models with different gas-to-dust ratios, scaled as in Figure \ref{fig:CRIRES_h2o_models}, compared to data between 2.912 and 2.945 $\mu$m.  Hatched regions mark OH emission. \label{fig:gtds}}
\end{figure*}

As shown in Figure \ref{fig:radlite_cpd}, our nominal RADLite disk model provides a reasonable match to the data.  Although we did not attempt exhaustive model fitting, the disk parameters in our nominal RADLite model --- high temperature, gas-to-dust ratio and turbulent broadening, combined with a very small emitting area, are analogous to the best-fit slab model parameters. Figure \ref{fig:gtds} shows how the appearance of the spectrum varies with different gas-to-dust ratios. In this figure, all models are scaled to the flux between 2.9292 and 2.9295 $\mu$m to investigate the shape of the spectrum, rather than the overall line flux. As the gas-to-dust ratio increases, the line spectrum ``flattens out'' as lines with low optical depth rise relative to lines with high optical depth.  Water abundance and gas-to-dust ratio can both affect the spectrum in this way, and cannot easily be disentangled --- the key conclusion that the observable water column must be large is the same as for the slab model.  

\begin{figure}
 \epsscale{1.0}
 \plotone{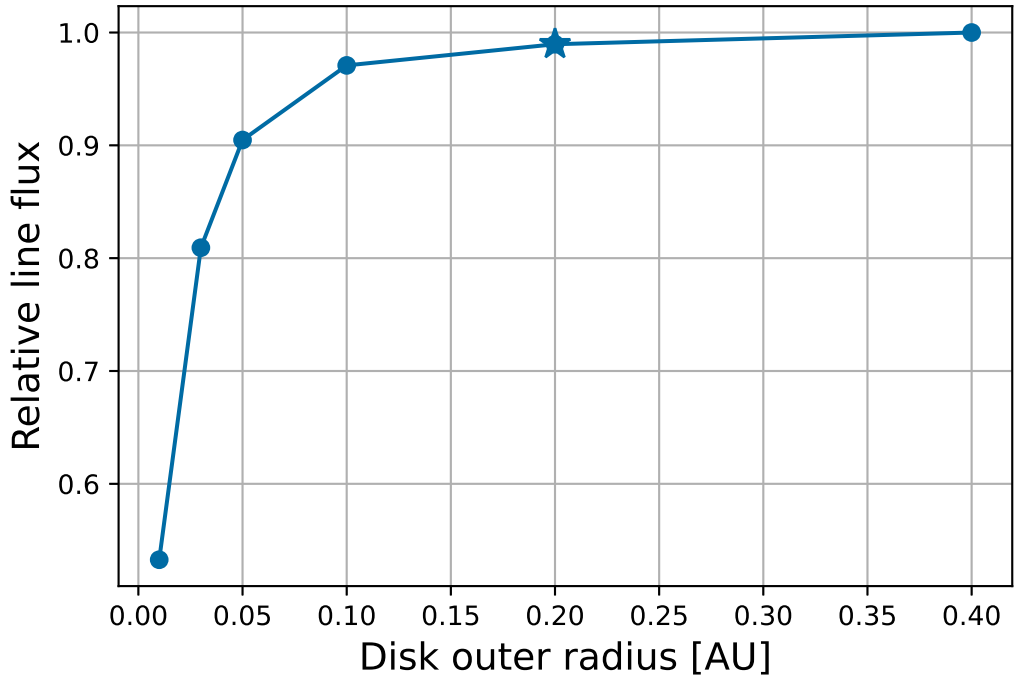}
 \caption{Variation in integrated line flux between 2.920 and 2.945 $\mu$m as a function of the disk outer radius. The line flux values are expressed relative to the line flux in the model with $R_{\rm disk} = 0.4\,$au. The star indicates the radius of the nominal RADLite model. \label{fig:lineflux_radius}}
\end{figure}

Figure \ref{fig:lineflux_radius} shows how the line flux in a portion of the spectrum (2.920--2.945 $\mu$m) changes as the disk outer radius (more precisely, the power law break location) is varied; we can see that $\sim$90\% of the flux in this wavelength region is contained within 0.05 AU of the planet. Therefore, our RADLite modeling also confirms the conclusion that the L-band emission comes from a small region.

In short, we find that it is possible to create a radiative transfer model of a CPD that recreates the observed $L$-band water spectrum, but the emitting region must be small and hot, with a high gas-to-dust ratio, high water vapor abundance, and high turbulent broadening. 

\subsection{Constraints from other data sets}
\subsubsection{Cooler water vapor from VV CrA A observed with Spitzer-IRS and VISIR}
\label{sec:spitzer_visir}

Spitzer-IRS spectra of the combined binary system show pure rotational emission from water vapor (Figure \ref{fig:spitzer}). The water lines at wavelengths longer than 12\,$\mu$m appear to be similar to those observed in many other protoplanetary disks, like AS 205, shown for comparison in Figure \ref{fig:spitzer} \citep{Pontoppidan10a}.  Since line fluxes rise more rapidly with wavelength for VV CrA than AS 205, the water reservoir for VV CrA may be cooler.  Similarly, the high-resolution 12\,$\mu$m VISIR spectra (Figure \ref{fig:VISIR_overview}), which resolve the binary, also show water vapor emission from VV CrA A in several rotational transitions. 

\begin{figure*}
\epsscale{1}
\plotone{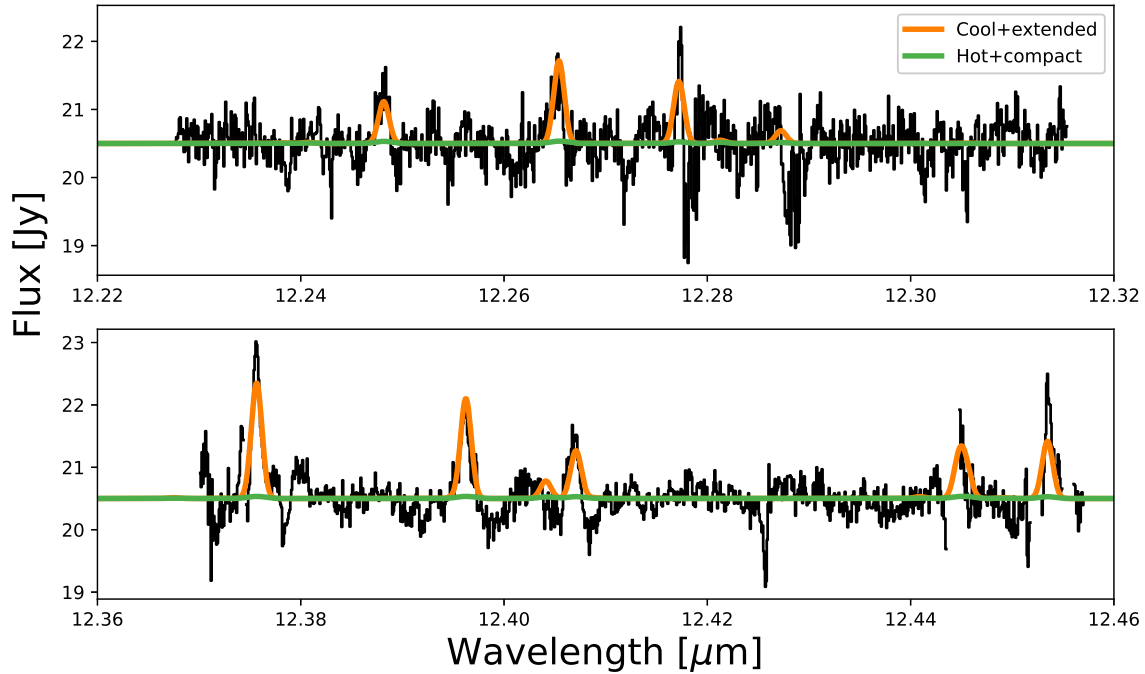}
\caption{VISIR spectra of VV CrA A in the ``OH'' and ``H$_2$O'' settings, along with two slab models.  Green: Our nominal $L$-band slab model extended to the $N$ band.  Orange: A slab model with $T=1000$ K, $A = 3$ AU$^2$ and $N_\mathrm{H2O}=10^{18}$ cm$^{-2}$.  \label{fig:cool_model}}
\end{figure*}

These rotational water lines appear to trace a different reservoir of water vapor than the ro-vibrational lines in the $L$ band. Because of the small emitting area and high temperature of the compact reservoir discussed in Section \ref{sec:water_reservoir}, our nominal $L$-band slab model, as well as our RADLite model, produce undetectably small line fluxes in VISIR and Spitzer-IRS observations. For example, assuming an $N$-band continuum flux of 20.5 Jy \citep{Sullivan19}, our nominal $L$-band slab model predicts a line-to-continuum ratio of $\sim0.2\%$ for the 12.3757 $\mu$m line observed with VISIR, in contrast to the observed ratio of $\sim$10\%.  We show this model in Figure \ref{fig:cool_model}.   Therefore, the unique water reservoir only appears in the short-wavelength spectra and does not appreciably affect the rotational water lines beyond 12\,$\mu$m.

Consequently, there must be an additional reservoir of cooler gas in the VV CrA A disk, possibly with larger emitting area and lower column density. Following the analysis described in \citet{Salyk19}, observed ortho/para line flux ratios slightly less than 3 imply that the $N$-band lines are borderline optically thick, and consistent with a column density near $10^{18}$\,cm$^{-2}$. Furthermore, we find that the data are consistent with a temperature of $\sim$1000 K and an emitting area of $\sim$3 AU$^2$, as shown in Figure \ref{fig:cool_model}.  This model, however, predicts flux in a few L-band emission features 30 times stronger than what we observe.  The extrapolation to the L band is highly temperature dependent, however, and lower temperature models are also consistent with the N-band model to within uncertainties; for example, a model with
 $T=600$ K, $N_\mathrm{H2O}=10^{18}$\,cm$^{-2}$ and $A=18$ AU$^2$ can reproduce the N-band data and is also consistent with the L-band data.  There may also be variability in the L- or N-band emission \citep[e.g.][]{Banzatti14,Banzatti15}, or variable continuum levels could affect our absolute flux calibration.

As seen in Figure \ref{fig:h2o_sa}, no SA signal is detected in our VISIR data. As discussed in \citet{Pontoppidan11}, the observed SA signal is affected by continuum dilution of the line emission, the effect of which is to make the observed signal smaller than the true line offset. The observed offset is reduced by a factor (1+F$_c$/F$_l$), where $F_c$ is the continuum flux and $F_l$ the line flux. The strongest upper limit on the spatial extent of the emitting area is therefore usually provided by the emission line with the largest line-to-continuum ratio. The line-to-continuum ratio for VV CrA A in the 12.3757 $\mu$m line is 0.11, corresponding to a continuum dilution factor of 10.1. Since the standard deviation of the SA signal is 0.7 AU (equivalent to 4 mas), we can constrain the  water emission for the 12.3757$\mu$m line to within a radius of $\lesssim$7 AU from the star.  

Overall, the VISIR and Spitzer spectra from VV CrA A are consistent with emission from warm ($\lesssim$1000 K) water vapor in the few AU region of the protoplanetary disk. The similarity of the Spitzer \citep{Pontoppidan10a} and VISIR \citep{Salyk19} spectra to other water-rich protoplanetary disks is also consistent with this hypothesis. Therefore, VV CrA A must harbor a cooler ($T\lesssim1000$ K), larger reservoir of water vapor in addition to the hot ($T\gtrsim1500$ K), compact reservoir discussed in Section \ref{sec:water_reservoir}.

\subsubsection{CO emission and constraints on the H$_2$O/CO ratio}
\label{sec:h2o_co_ratio}
While high water column density is required to fit the $L$-band observations, we do not know if the spectra arise from a region with high overall gas column or an enhanced H$_2$O abundance relative to other molecular species. By analyzing the CO emission in the $K$ and $M$ bands, it may be possible to distinguish between these two scenarios.

\begin{figure*}
 \epsscale{1.3}
 \plotone{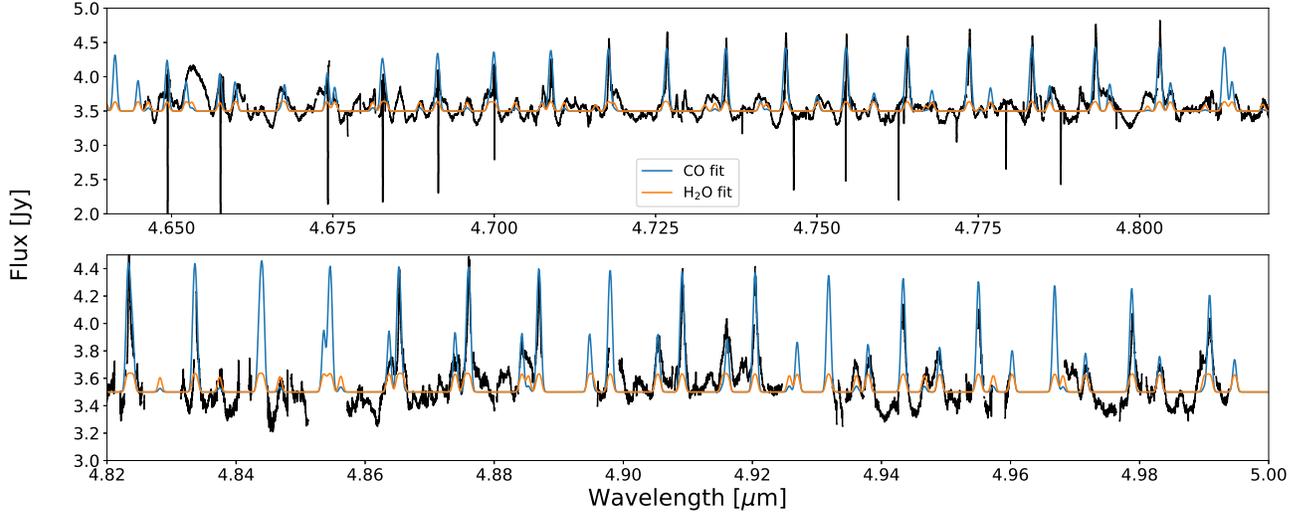}
 \caption{M-band CRIRES spectrum of V VCrA A with best-fitting CO slab model ($T=1000\ $K, $N_\mathrm{CO}=1\times10^{19}\ \mathrm{cm}^{-2}$, $v=10\ \mathrm{km\ s}^{-1}$ and $A = 0.06$ AU$^2$; blue) and a model with the same parameters as our nominal L-band model ($T=1500\ $K, $N_\mathrm{CO}=3\times10^{20}\ \mathrm{cm}^{-2}$, $v=10\ \mathrm{km\ s}^{-1}$ and $A = 0.003$ AU$^2$;
 orange), assuming $\rm H_2O:CO=1$.  \label{fig:CRIRES_modeling_mband}}
\end{figure*}

Figure \ref{fig:CRIRES_modeling_mband} shows the CO $M$-band spectrum for VV CrA A along with a slab model fit to these $M$-band data, found by exploring the same range of temperature, column density, and turbulent broadening as for the $L$-band water emission. We find that the CO $M$-band emission is better fit with a lower temperature ($T\sim1000$ K) and lower column density ($N_\mathrm{CO}\sim10^{19}\,\rm cm^{-2}$) than our best fit to the $L$-band water emission. If we instead take our nominal compact, hot {\it L-band model}, set H$_2$O:CO=1, and predict the $M$-band CO emission, we find line fluxes smaller than observed. Therefore, the $M$-band CO data seem to probe a cooler, larger, reservoir than the $L$-band data. At the same time, they are fully consistent with the presence of an underlying compact, hot region.

\begin{figure*}
 \epsscale{1.3}
 \plotone{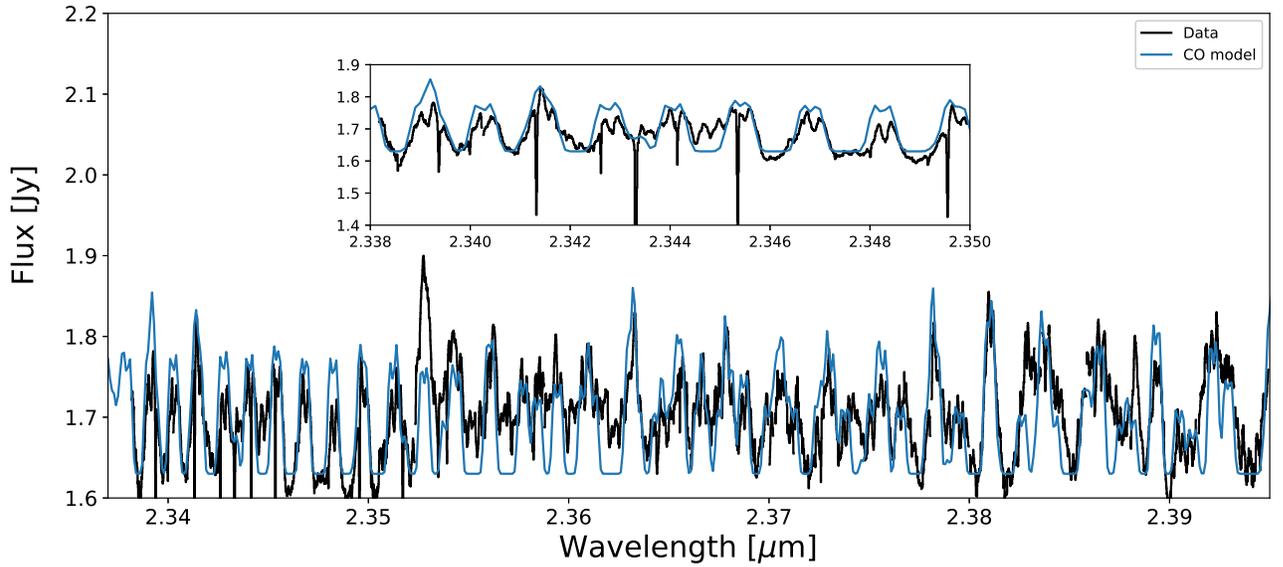}
 \caption{CRIRES K-band spectrum of VV CrA A (black) and a slab model convolved with the inner disk line profile (Figure \ref{fig:lineshape_cartoon}, left panel) discussed in Section \ref{sec:lineshape_models} (blue).  The inset highlights the double-peaked line profiles \label{fig:kband_modeling}}
\end{figure*}

The CRIRES $K$-band spectrum (Figure \ref{fig:kband_modeling}) shows a large number of CO overtone (v=2-0, 3-1 and 4-2) emission features.   A CO slab model is shown, in which the emitting area is constrained to be the same as that of the nominal $L$-band model. There are a few water vapor transitions in this region, but it is not entirely clear whether the addition of a water model improves the fit to the data.  As an example, we note the mismatch between data and CO model in the 2.383--2.385 $\mu$m region, which may indicate the presence of water vapor. The temperature in the CO model is around 2000 K and column density is very high at $N_\mathrm{CO}\sim5\times10^{20}\ \mathrm{cm}^{-2}$. However, most lines are optically thin or only marginally optically thick, so the best-fit CO column density is dependent on our assumed emitting area.  If it is assumed that the K-band CO and L-band water probe the same reservoir, this implies a column density ratio of $\rm H_2O/CO \sim 0.7$ --- slightly lower than typical column density ratios of $\rm H_2O/CO \sim 1$ measured in inner disks \citep{Salyk11}. However, as we discuss in Section \ref{sec:lineshapes}, line shapes suggest these may not necessarily be probing the same reservoir.  As shown in the inset of Figure \ref{fig:kband_modeling}, the K-band CO lines are double peaked. In fact, the shape is well fit by the inner disk ring model discussed in \ref{sec:lineshape_models}, which is located at the dust inner radius of 0.17 AU.

\subsubsection{Velocity information}
\label{sec:lineshapes}

\begin{figure}[ht!]
\epsscale{1}
\plotone{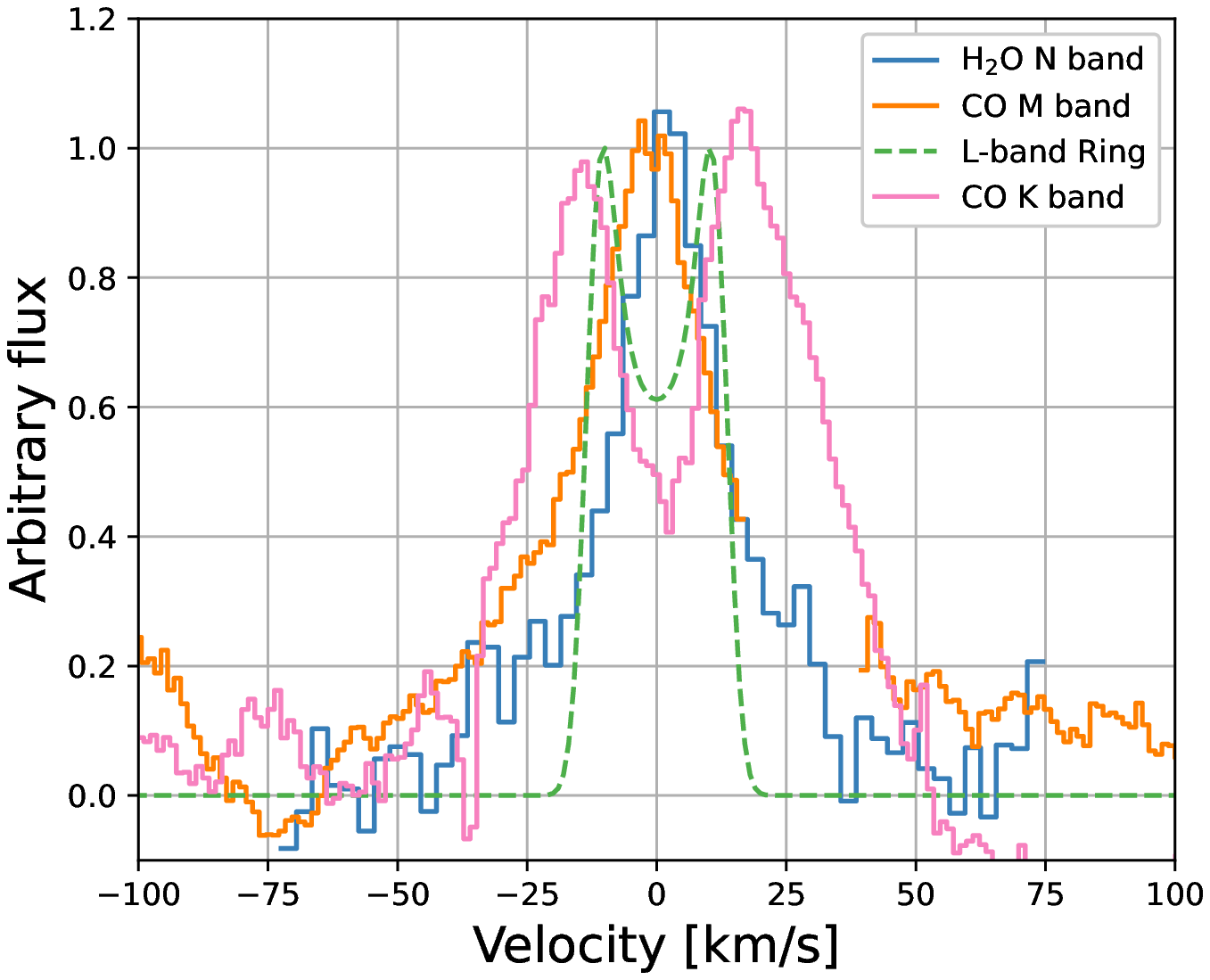}
\caption{A comparison of observed line shapes: VISIR water vapor line composite (blue), CO M band (v=1-0 P(22); orange), CO K band (v=3-1 R(15); pink), and our L-band wide ring model (green, dashed --- see Section \ref{sec:lineshape_models}).  Velocities are stellocentric, assuming a stellar velocity of $-1$ km s$^{-1}$ \citep{Fang18}. \label{fig:combined_lineshape}}
\end{figure}

Emission lines from each wavelength regime appear to have distinct line shapes, suggesting they each probe different reservoirs.  We show a compendium of line shapes in Figure \ref{fig:combined_lineshape}.  The CO K-band spectrum is double peaked, with peak locations near 28 km s$^{-1}$ --- similar to the inner disk model shown in Figure \ref{fig:lineshape_cartoon}.  VISIR and CO M-band lines appear Lorentzian in shape, with M-band lines being broader.  It is not possible to extract an isolated H$_2$O line from the L-band spectrum, so we show here our wide ring model (middle  panel of Figure \ref{fig:lineshape_cartoon}). This particular model appears intermediate in width between the K- and M-band lines, although it is not uniquely specified by the L-band data; the CPD model, for example, is narrower than the M-band line shape.

 A simple understanding of protoplanetary disks would predict narrowing of line profiles as observations probe longer wavelengths (i.e., as one probes from the K to the N bands), although it should be kept in mind that emission lines at similar wavelengths may probe a variety of excitation levels (see, for example, Figure \ref{fig:model_breakdown}).  The K-, M-, and N-band data follow this trend --- narrowing with increasing wavelength, and the K-band data appears well-described with a line shape produced at the dust inner rim (Figure \ref{fig:kband_modeling}). Non-Keplerian motion is likely required to explain the Lorentzian-like M- and N-band profiles \citep[e.g.][]{Bast11}.  Lorentzian-like M-band CO rovibrational emission has previously been attributed to the presence of low-velocity disk winds \citep{Pontoppidan11, Banzatti22}.  A contribution from a CPD would also cause deviations from the simple disk velocity trend, but we cannot say definitively how the L-band emission line profile compares to other wavelengths.

The plot also shows that emission velocities all appear close to the stellar velocity.   Although in this plot the M-band CO appears shifted by $\sim$3 km s$^{-1}$, analysis of a suite of emission lines suggests no statistically significant shift, with a systematic uncertainty of a few km s$^{-1}$.  Similar uncertainties of a few km s$^{-1}$ are found for the K- and N-band lines.  In short, we find no statistically significant deviations of any velocities from the systemic velocity.

\subsection{VV CrA B}
\begin{figure*}
 \epsscale{1.0}
 \plotone{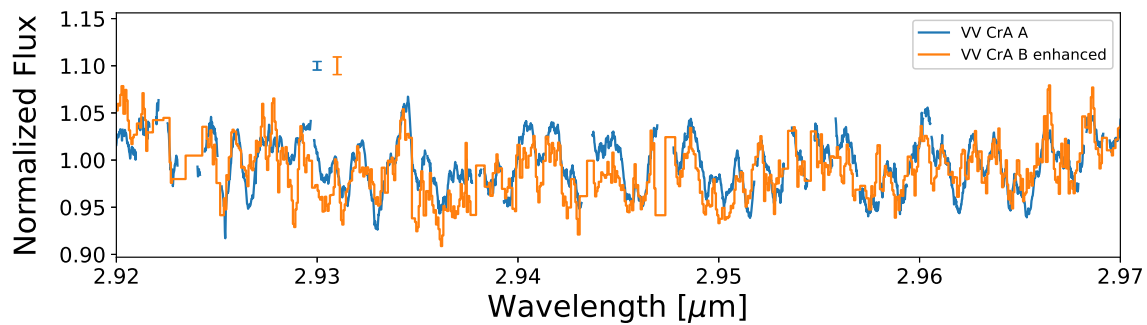}
 \caption{Comparison between portions of the L-band spectra of VV CrA A and an enhanced spectrum of VV CrA B (see text for details). Representative error bars (assumed to be 3 $\times$ the statistical error) are shown in the upper left.  \label{fig:H2O_AB_comparison}}
\end{figure*}

The $L$-band spectrum of VV CrA B has a lower SNR and lower line/continuum ratios than the spectrum of VV CrA, but shows some of the same features.  In Figure \ref{fig:H2O_AB_comparison}, we show a comparison between the VV CrA A spectrum and the spectrum of VV CrA B modified in two ways --- it has been binned by a factor of 5, and then had its line/continuum ratio enhanced by a factor of 3.  With these enhancements, there is a strong correspondence between the shapes of two spectra --- the data sets shown in this figure have a Pearson R coefficient of 0.96.

\begin{figure}
\epsscale{0.5}
\plotone{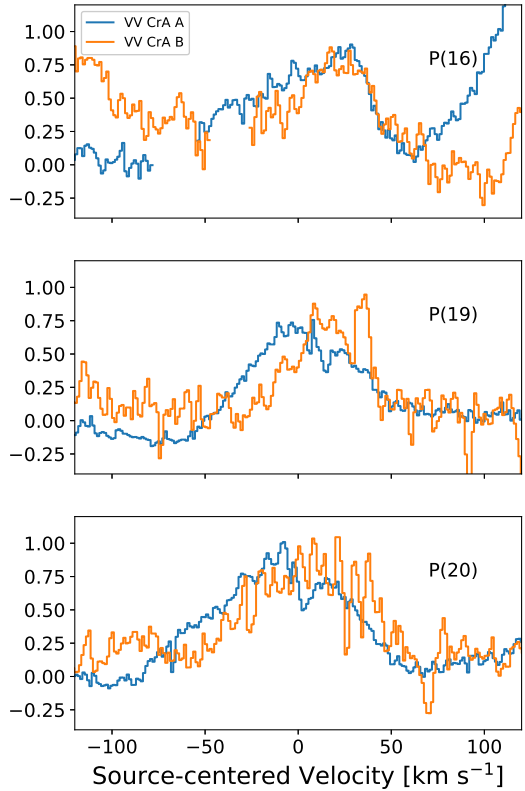}
\caption{Normalized CO v=2-1 emission from VV CrA A (blue) and VV CrA B (orange). \label{fig:co_lineshapes_AB}}
\end{figure}

The M-band spectra from CRIRES (Figure \ref{fig:crires_overview}) reveal primarily deep CO absorption lines from VV CrA B, as discussed in \citet{Smith09}, although three CO $v=2-1$ lines are detected in emission --- see Figure \ref{fig:co_lineshapes_AB}.  A comparison of the emission from the two stars in the same transitions shown in Figure \ref{fig:co_lineshapes_AB} reveals similar emission line shapes for the two disks, suggesting similar CO emitting locations and disk inclinations.   A slight velocity offset is observed, but the SNR of the VV CrA B data is too low to make this result significant.

\begin{figure*}
  \centering 
  \epsscale{1}
\plotone{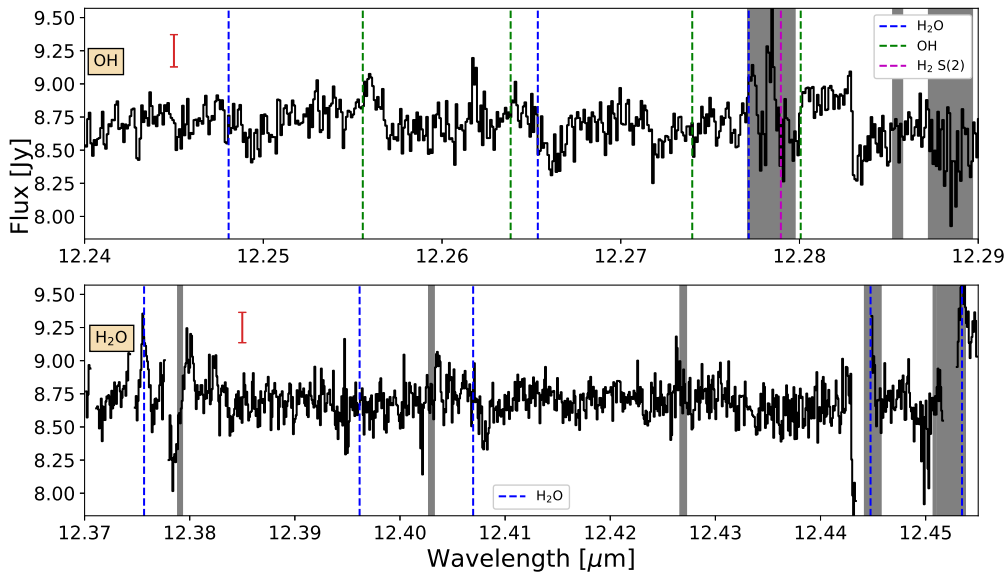}
\caption{Reduced spectra of VV CrA B in the two water/OH settings observed with VISIR.  Data wavelengths are heliocentric and theoretical wavelengths have been shifted by $-$21 km s$^{-1}$, the velocity of CO $v=2-1$ emission from VV CrA B.  In the upper left, we show representative error bars computed from the standard deviation of flat spectral regions. \label{fig:VISIR_B} }
\end{figure*}

There is a marginal detection of water vapor in VV CrA B at 12.3757 $\mu$m in the combined H$_2$O VISIR setting (Figure \ref{fig:VISIR_B}) and at multiple transitions for the 2017 Sep 08 data alone (Figure \ref{fig:h2o_sa}).  The difference in the water line-to-continuum ratio for the two disks could arise from differences in excitation conditions (temperature, density, etc.), different water abundances, or different inclination angles. However, the similarity of the CO emission line shapes in the two disks argues against different inclinations.

In short, the spectra of VV CrA B suggest that similar reservoirs of water vapor and CO exist in the disks around VV CrA A and B, albeit with some difference causing a lower line/continuum ratio for B.  Higher SNR data would be useful to confirm this result.  The similar CO emission line shapes for the two disks also support a scenario in which the CO absorption seen towards VV CrA B is caused by foreground absorption by VV CrA A, rather than self-absorption in an inclined disk around VV CrA B.

\section{Discussion}
\label{sec:discussion}
\subsection{Correlation with evolutionary state}

To our knowledge, the complex L-band spectral patterns we present in this work have only been seen in one prior study, of V1331 Cyg \citep{Doppmann11}.  A possible reason for this rarity may be that most of the high-resolution L-band spectroscopic studies have focused on later evolutionary states \citep{Salyk08,Mandell12,Banzatti17}. The age of VV CrA is somewhat uncertain, with estimates ranging from a fraction of a Myr \citep{Prato97} to several Myr \citep{Scicluna16}.  But the accretion rate of VV CrA A ($3\times10^{-7} M_\odot\,\mathrm{yr}^{-1}$; \citealp{Fang18}) is higher than that of typical T Tauri stars, and it is associated with a Herbig-Haro object \citep{Wang04}, indicative of an outflow.  The presence of similar water emission features in the coeval disks around VV CrA A and B is also consistent with this reservoir correlating with evolutionary state. 

The presence of CO overtone emission from VV CrA A also makes it an atypical T Tauri star.  CO overtone emission is more commonly observed around hotter (especially B-type) Herbig stars \citep{Ilee14}.  It has been detected from some T Tauri stars as well, but is associated with outflows \citep{Carr89}.  Given the high temperatures required to excite the CO overtone lines, their presence from a T Tauri star may require an accretion burst to provide additional localized heating \citep{Ilee14}.  The strongly single-peaked emission lines observed from VV CrA A in the M band are also associated with low-velocity molecular disk winds \citep{Pontoppidan11, Bast11, Pontoppidan22}, which may be stronger when the star is more actively accreting.  

A similar reservoir of both CO and water with high gas column densities, high temperatures, and high turbulent velocities  was observed around SVS 13 \citep{Carr04}, a young binary system \citep{Diaz-Rodriguez22} in NGC 1333 \citep{Strom76}.  SVS 13 is a Class I YSO \citep{Bitner08} with higher than average accretion ($6\times 10^{-7}$ M$_\odot$ yr$^{-1}$; \citealp{Edwards03}),  and both outflow and wind signatures \citep[e.g.][]{Edwards03,Diaz-Rodriguez22}.  Unfortunately, no direct comparison can be made with our spectra because the two sources were not observed at the same wavelengths. 

A direct comparison can, however, be made with an L-band spectrum of V1331 Cyg, presented by \citet{Doppmann11}.  Similar to our work, they show that the many water features in the V1331 Cyg spectrum have similar line strengths, requiring a high column density.  They report best-fit water model parameters similar to ours --- $T=1500$ K, $N_\mathrm{H2O}=2\times10^{21} $cm$^{-2}$ and $A=0.02$ AU$^2$ --- albeit with somewhat larger water vapor column density and emitting area and no additional turbulent broadening.  V1331 Cyg is a young A8-G5 star in L988, with an accretion rate $>3\times10^{-7}$ M$_\odot$ yr$^{-1}$ \citep[see][and references therein]{Doppmann11}.

Does a correlation with evolutionary state, and with the presence of outflows and/or winds, help constrain the physical origin of the emission?  We might expect these properties to be associated with high inner disk gas temperatures and turbulent velocities, just as observed in the L-band data.  In addition, it is plausible that a high accretion luminosity could push dust sublimation outwards, while water self-shielding \citep{Bethell09} maintains a region of high water abundance close to the star; in combination, this could produce a region with high observable water columns, as seen in our L-band data.  In short, VV CrA's evolutionary state provides plausible ways to produce conditions in the inner disk consistent with those seen in the L-band data.  

As discussed in Section \ref{sec:lineshape_models}, there is not a good match between observed and predicted lineshapes for an inner disk model.  However, given VV CrA's evolutionary state and single-peaked CO emission lines, it is likely that it has a molecular wind \citep[e.g.][]{Bast11,Pontoppidan11,Banzatti22}.  Perhaps the non-Keplerian motions in this wind could alter water line shapes sufficiently to reconcile the observed spectral shape with gas located close to the dust inner rim (i.e., the inner ring model of Figure \ref{fig:lineshape_cartoon}).  In other words, the emission could arise not only from the disk itself, but from the base of the molecular wind.  A higher velocity or extended outflow origin, however, appears inconsistent with the low (same as systemic) line center velocities and small spatial extent of the emission.

As for the CPD hypothesis, the fact that the observed hot water reservoir appears associated with early evolutionary states would imply that the CPD must be easier to detect in early stages than later stages of disk evolution.  This could naturally be explained by a reduction in planetary luminosity and/or CPD viscous heating with time, or by changes in H$_2$O abundances.  CPD models do predict variations in both H$_2$O abundances \citep[e.g.][]{Ilee17} and temperatures \citep[e.g.][]{Canup02} with time that could potentially create distinct observability windows.  In addition, the models of \citet{Marley07} predict a rise in planetary luminosity during a brief period between 2--3 Myr.

It may be argued that if CPDs are rare, it would be unlikely for one to detect similar signatures from both VV CrA A and B.  However,  dynamical signatures of protoplanets are being detected in a majority of bright disks observed with ALMA at high resolution \citep[e.g.][]{Andrews18,Pinte20}, and binarity only appears to affect planetary occurrence rates for small separations \citep[e.g.,][]{Kraus16}.  Therefore, it may be reasonable for CPDs to exist in both disks of the binary pair.

\subsection{Is our CPD model realistic?}
Is our RADLite model in Section \ref{sec:radlite_model} a realistic model of a CPD?  We have very few observational constraints on CPD properties, but can compare to the observed disk around PDS 70c \citep{Isella19}, and existing CPD models.  The PDS 70c CPD has a dust mass between 2$\times10^{-3}$ and $4.2\times10^{-3}$; the gas mass is unconstrained. Our model's high gas-to-dust ratio of 70,000 implies a dust mass of only 2.7$\times10^{-5}$ M$_\oplus$ --- significantly lower than that of PDS 70c's CPD.  In addition, our model disk size is 0.2 AU, while that of PDS 70c's CPD is estimated to be between 1.4 and 3 AU.  However, models suggest that CPD size should be proportional to the Hill radius \citep[e.g.][]{Ayliffe09,Quillen98} and thus, semi-major axis.  Therefore, our model disk could correspond to a CPD 7--15 times closer than the 35 AU orbital radius of PDS 70c, i.e., a disk at  2.3--5 AU. A smaller CPD might also have a lower overall dust mass, again, consistent with our model.  In addition, as shown in Figure \ref{fig:lineflux_radius}, our model is not sensitive to emission beyond $\sim0.05$ AU, so a larger disk would also be consistent with the observed spectrum.  

PDS 70c is estimated to have a mass between 4 and 12 M$_J$, consistent with our modeled planet mass of 6 M$_J$. However, the luminosity of PDS 70c is estimated to be $\sim6\times10^{-5}\,L_\odot$, while our model planet has a significantly higher luminosity of $0.02\, L_\odot$.  Our model essentially requires a high luminosity to produce the strong observed water line fluxes, as the model assumes the disk is passively heated.  While different from PDS 70c's luminosity, our model's luminosity is fully consistent with core accretion evolutionary tracks in which the planetary luminosity rises higher than 0.02 $L_\odot$ during runaway gas accretion \citep{Marley07}.  As an alternative to passive heating from the central planet, the CPD might also be actively heated by accretion from within the disk itself.  This could be consistent with the high turbulent broadening required to match the observed spectra.  Indeed, models predict temperatures consistent with those we observe \citep[e.g.][]{Ayliffe09,Zhu15} in at least parts of the CPD.  In addition, models predict enhanced H$_2$O abundances as the high temperatures cause sublimation of icy grains \citep[e.g.][]{Ilee17,Ayliffe09}.

We might also compare to the solar system's Galilean satellites.  Our modeled water emitting area is 0.003 AU$^2$, while a disk extending to Callisto's semi-major axis would have an area about a factor of 6 smaller --- $5\times10^{-4}$AU$^2$.  As discussed in \citet{Canup02}, the planetary accretion can potentially raise local temperatures several thousands of degrees.  In addition, the formation of the Galilean satellites may have required a continual enhancement of metallicity in the inner disk; these solids would sublime at high temperatures, potentially resulting in a disk with enhanced water-to-dust ratios at some stages of the disk's evolution. The high water column density seen in our L-band spectra could arise from such a water abundance enhancement.

\subsection{Distinguishing between models}
How might we use future observations to better understand the origin of the L-band water emission? It would be helpful to better understand the occurrence rate of this phenomenon, and, especially, whether its occurrence correlates with evolutionary state or any other disk properties.  Spectro-astrometric data in the M- and L-bands would also be helpful for characterizing any contributions to the emission from a molecular wind \citep{Pontoppidan11}.  If present in many young disks and if no ancillary data supports the existence of a CPD, it will be necessary to reconcile the observed line kinematics with our expectations for protoplanetary disks.  

On the other hand, a CPD might have additional detectable signatures.  It has been predicted for some time that CPDs would produce both asymmetries and variability in near-IR CO emission lines \citep{Regaly14}. Time variations and asymmetries have been seen in the spectra of the disk around HD 100546 along with variable spectro-astrometric signatures, and \citet{Brittain14} argue that these are consistent with the presence of a CPD.  Spectro-astrometry might be performed for VV CrA A in the L band, although with line-to-continuum ratios of 5--10\% for the water lines, it suffers from continuum dilution factors of 11--21.  With a spectro-astrometric RMS of 5 mas, one can detect separations of 8--16 AU.  The CPD would also have a central velocity that differs from that of the star, and varies throughout the planet's orbit.  The CPD's speed should be resolvable with CRIRES, but would require understanding and accounting for systematic uncertainties in our velocity measurements.  If the planet is at 2.3--5 AU, we would expect velocities of $\sim$ a few km s$^{-1}$, varying with a period of 4.7--15 years.  A spatial offset for the hot water lines could also be directly measured with an instrument like JWST's NIRSpec IFU, but unfortunately VV CrA A's brightness is above NIRSpec's saturation limits.

A CPD could also be searched for via direct imaging of its mm-wave emission \citep[e.g.][]{Zhu18} or accretion signatures \citep{Sallum15}, although its detectability would depend on its angular separation from the star.  Published mm-wave images \citep{Scicluna16} have a beam size of 1-2$''$, equivalent to $\sim$150-300 AU at 157 pc, so insufficient to detect a close-in embedded planet.  SEDs can also indicate large gaps, but the VV CrA disks are heavily embedded, impeding detections of any gaps \citep{Kruger11,Scicluna16,Sullivan19}.

Flipping this discussion on its head, this investigation suggests that it is at least plausible that expected CPD properties would produce a detectable L-band water emission signature.  Therefore, it would be worthwhile to couple hydrodynamic CPD models with radiative transfer to make more realistic predictions for the L band.  This could potentially provide a new avenue for discovering CPDs.

\section{Conclusions}
In this work, we have presented spectra of the VV CrA binary observed with VLT-CRIRES, VLT-VISIR, and Spitzer-IRS.  We find that the VV CrA A disk presents a water emission spectrum at mid-infrared wavelengths that is typical for T Tauri stars; in contrast, in the L-band, the spectrum contains an unusual number of high temperature lines with similar line strengths.  Using both slab models and the LTE disk modeling code RADLite, we show that producing such a spectrum requires a small reservoir of hot gas with a large water vapor column density and high turbulent broadening.  Line kinematics do not match well with a model in which the L-band emission arises from just outside the dust inner rim in a Keplerian disk. Since the presence of this reservoir is seemingly linked to the presence of high accretion rates and winds/outflows, non-Keplerian motion may be required to explain the emission. We also show that a compact spot, such as a CPD, may provide emission features consistent with those observed. Additional exploration of this phenomenon in other targets, exploration of variability and asymmetry of the molecular emission from VV CrA, and ancillary data to search for any possible CPD signatures, are warranted to better understand the physical origin of this unusual spectral pattern.

\acknowledgments
C.S. would like to thank all who assisted with the planning and observations in the VISIR large programme, including the support staff at the VLT. C.S. would also like to acknowledge helpful comments on this work from Michael R.\ Meyer and Greg  Doppmann.  C.H. is a former Winton Fellow and this research has been supported by Winton Philanthropies / The David and Claudia Harding Foundation. C.H. received funding from the European Unions Horizon 2020 research and innovation programme under the Marie Sklodowska-Curie grant agreement No 823823 (RISE DUSTBUSTERS project). C.H. and R.A. acknowledge support from the European Research Council (ERC) under the European Union’s Horizon 2020 research and innovation programme (grant agreement No 681601).
A.C. acknowledges funding from ANR of France under contract number ANR\-18\-CE31\-0019 (SPlaSH).
This work is supported by the French National Research Agency in the framework of the Investissements d'Avenir program (ANR-15-IDEX-02),
through the funding of the ``Origin of Life'' project of the Grenoble-Alpes University.

\appendix
\renewcommand\thefigure{\thesection.\arabic{figure}}    
\section{Complete reduced spectra}
\setcounter{figure}{0}  

\begin{figure*}
 \epsscale{1}
 \plotone{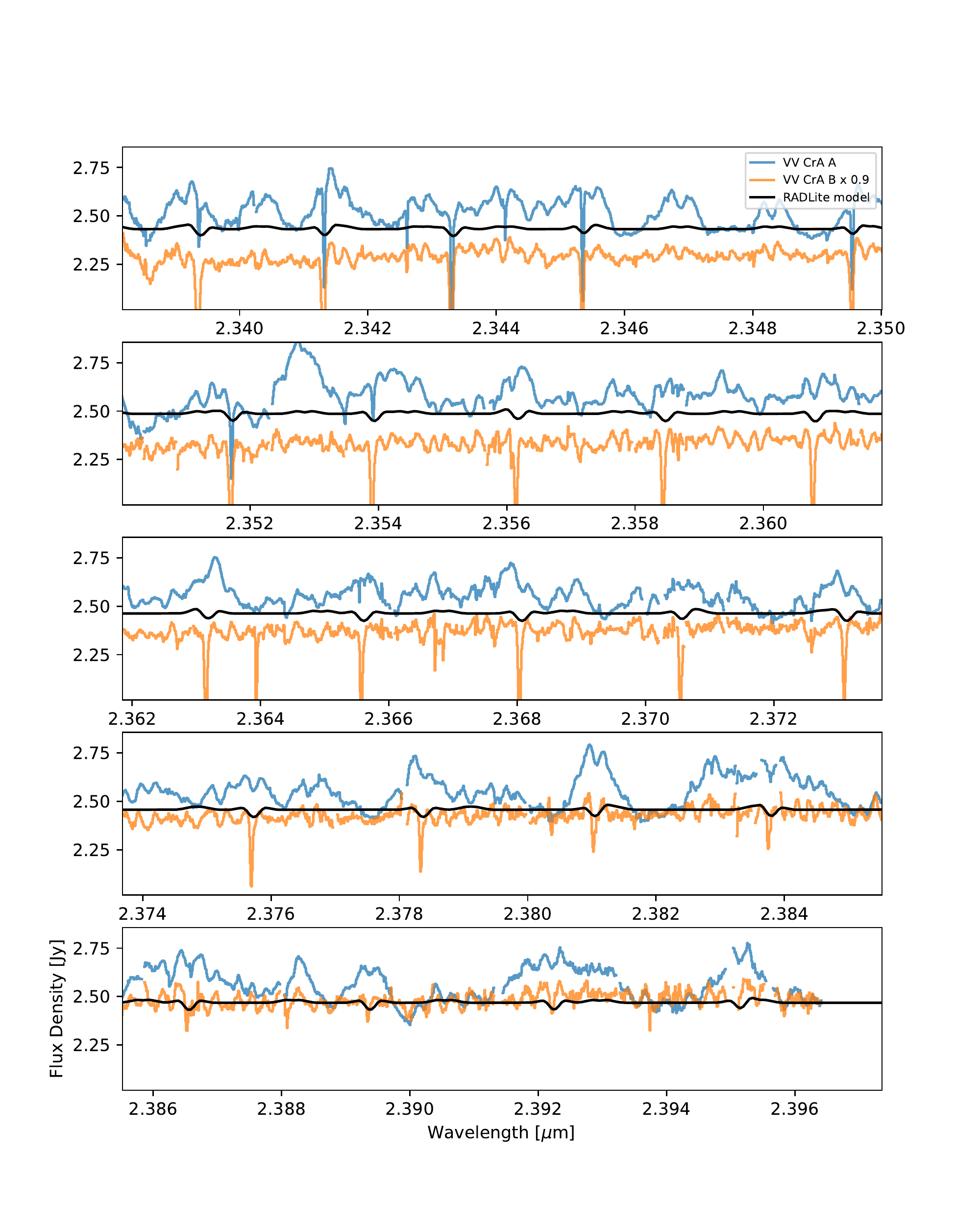}
 \caption{Reduced CRIRES spectra of VVCrA A (blue) and B (orange) in the K band.  In this and Figures \ref{fig:crires_A_H2O}--\ref{fig:crires_A_M}, the black curve shows our nominal L-band RADLite model. \label{fig:crires_A_K}}
\end{figure*}

\begin{figure*}
 \epsscale{1}
\begin{center}
 \plotone{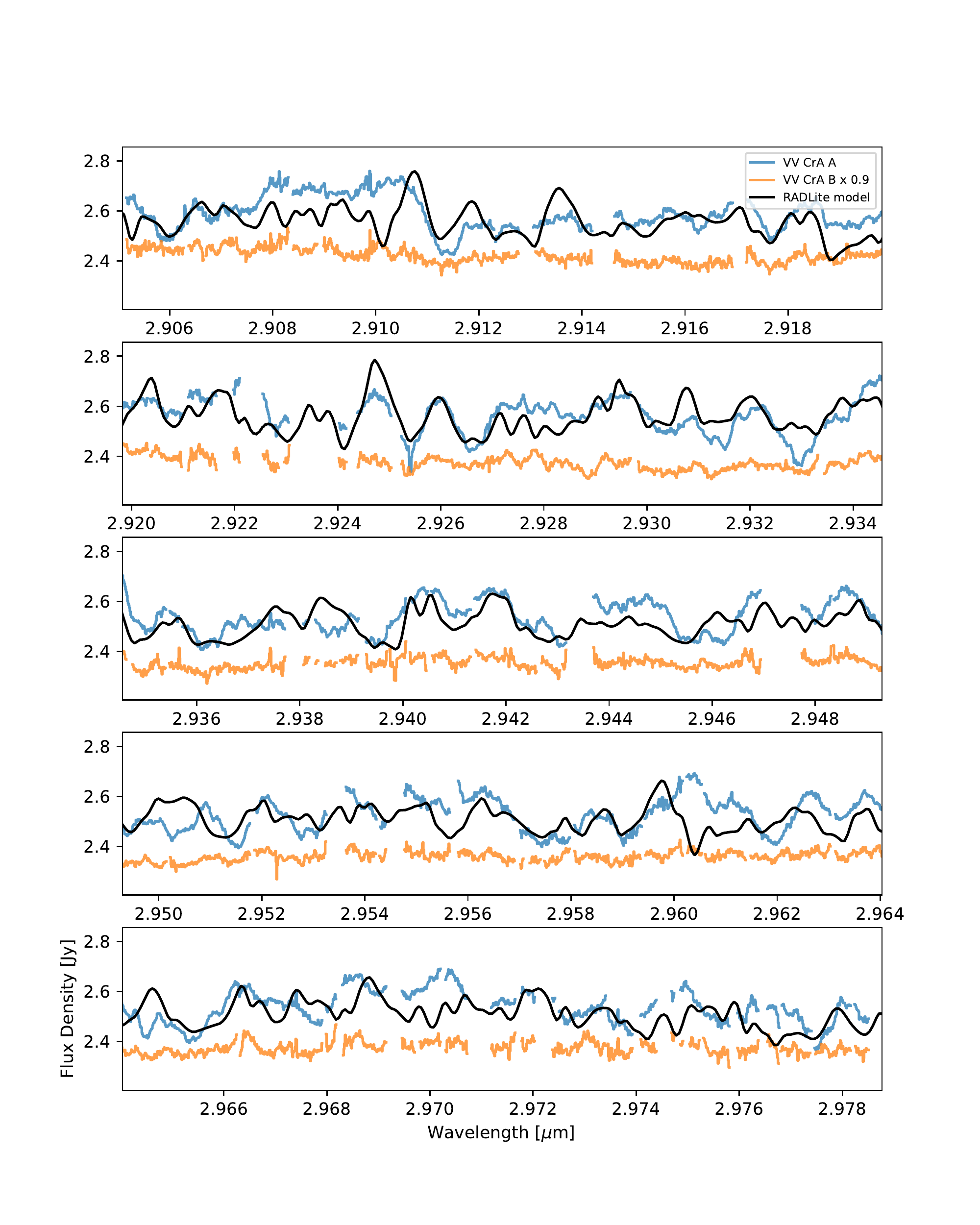}
\end{center}
 \caption{Reduced CRIRES spectra of VVCrA A and B in the L-band ``H$_2$O'' setting.  \label{fig:crires_A_H2O}}
\end{figure*}

\begin{figure*}
 \epsscale{1}
 \plotone{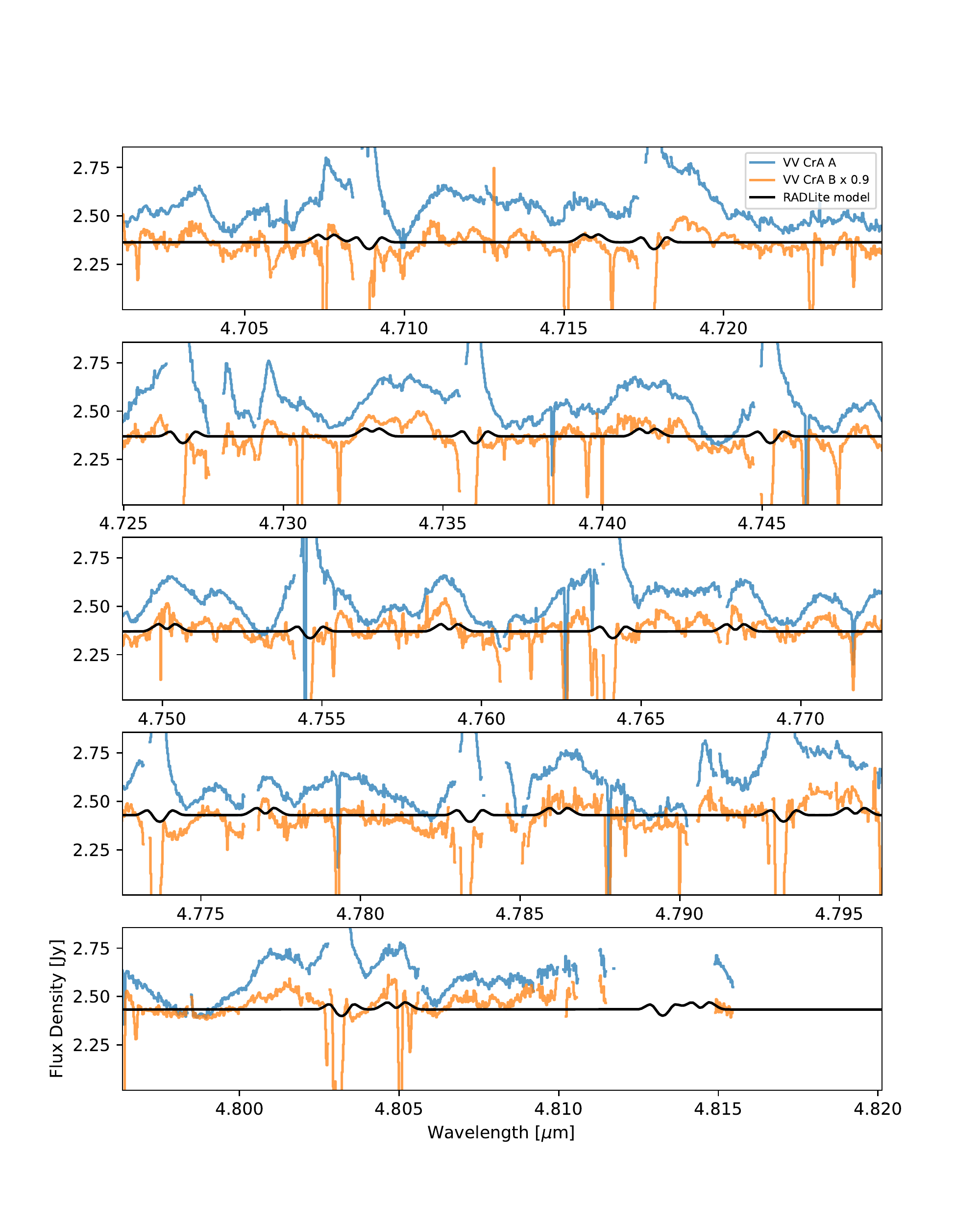}
 \caption{Reduced CRIRES spectra of VV CrA A and B in the M band.  \label{fig:crires_A_M}}
\end{figure*}

\begin{figure*}
\epsscale{1}
\plotone{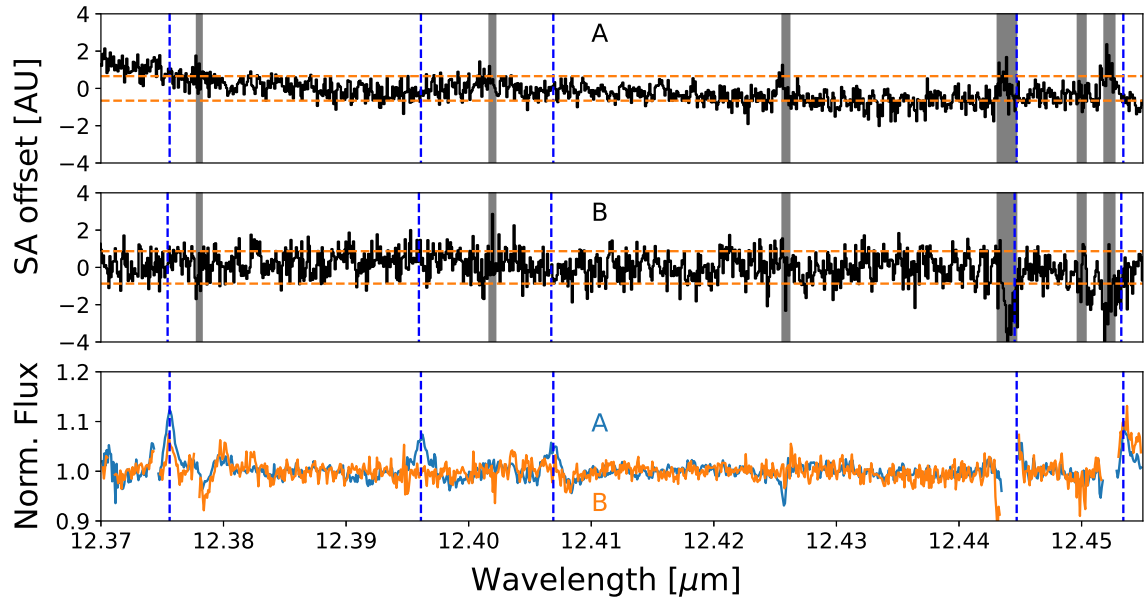}
\caption{Top two panels: Spectro-astrometric measurement in the 12.414 $\mu$m water setting for VV CrA A (top) and B (middle). Dashed horizontal lines mark plus and minus one standard deviation.  Dashed vertical lines mark theoretical locations of water transitions.  Bottom panel: The reduced spectra for VV CrA A (blue) and B (orange), for reference. \label{fig:h2o_sa}}
\end{figure*}

Figures \ref{fig:crires_A_K}--\ref{fig:crires_A_M} show complete CRIRES spectra of VV CrA in the $K$,$L$ and $M$ bands, along with our nominal RADLite CPD model.  Figure \ref{fig:h2o_sa} shows our VISIR spectro-astrometric data.

\begin{figure*}
 \epsscale{1}
 \plotone{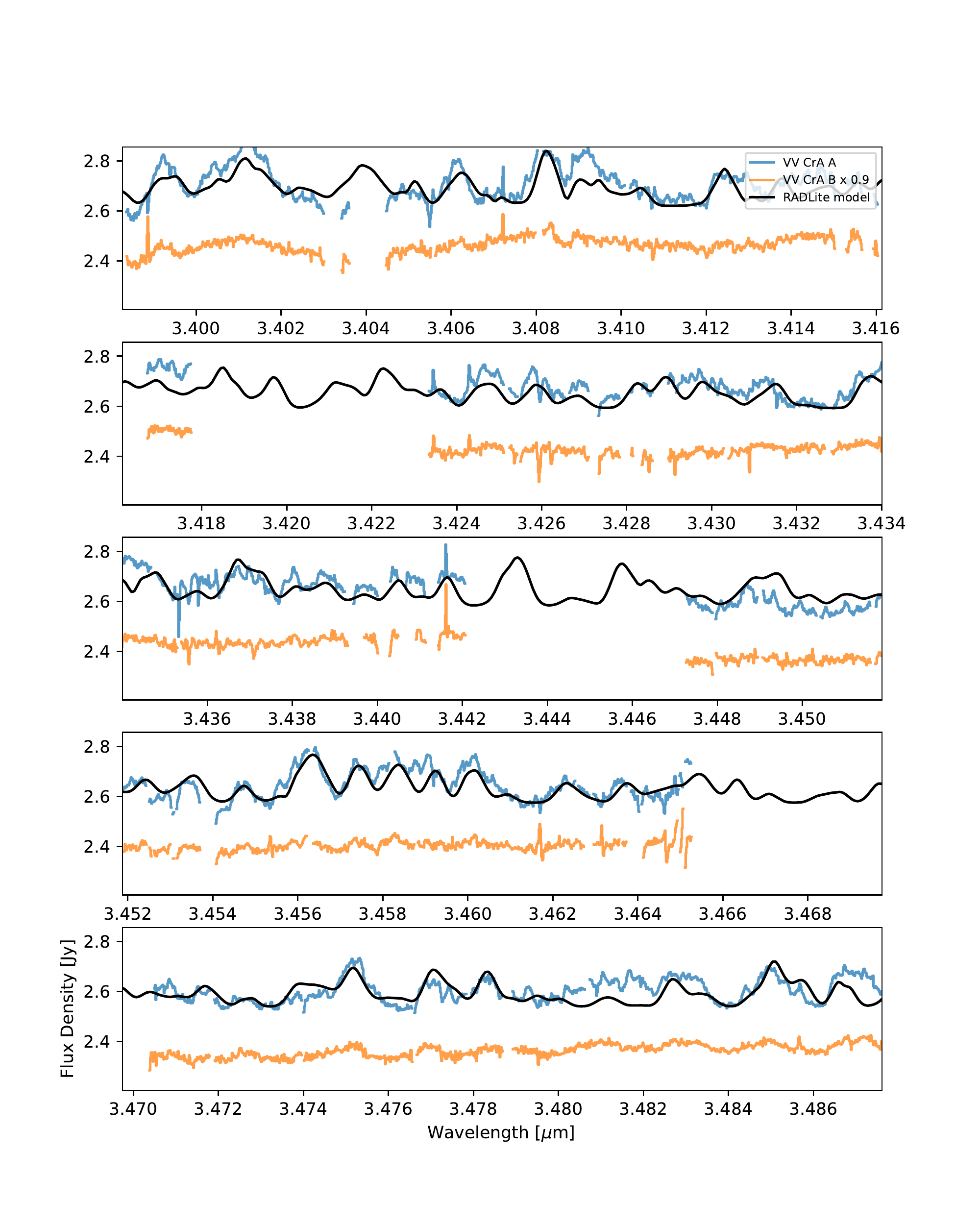}
 \caption{(a) First part of reduced CRIRES spectra of VV CrA A and B in the L-band ``organics'' setting.\label{fig:crires_A_org}}
\end{figure*}

\begin{figure*}\ContinuedFloat
  \centering 
  \epsscale{1}
\plotone{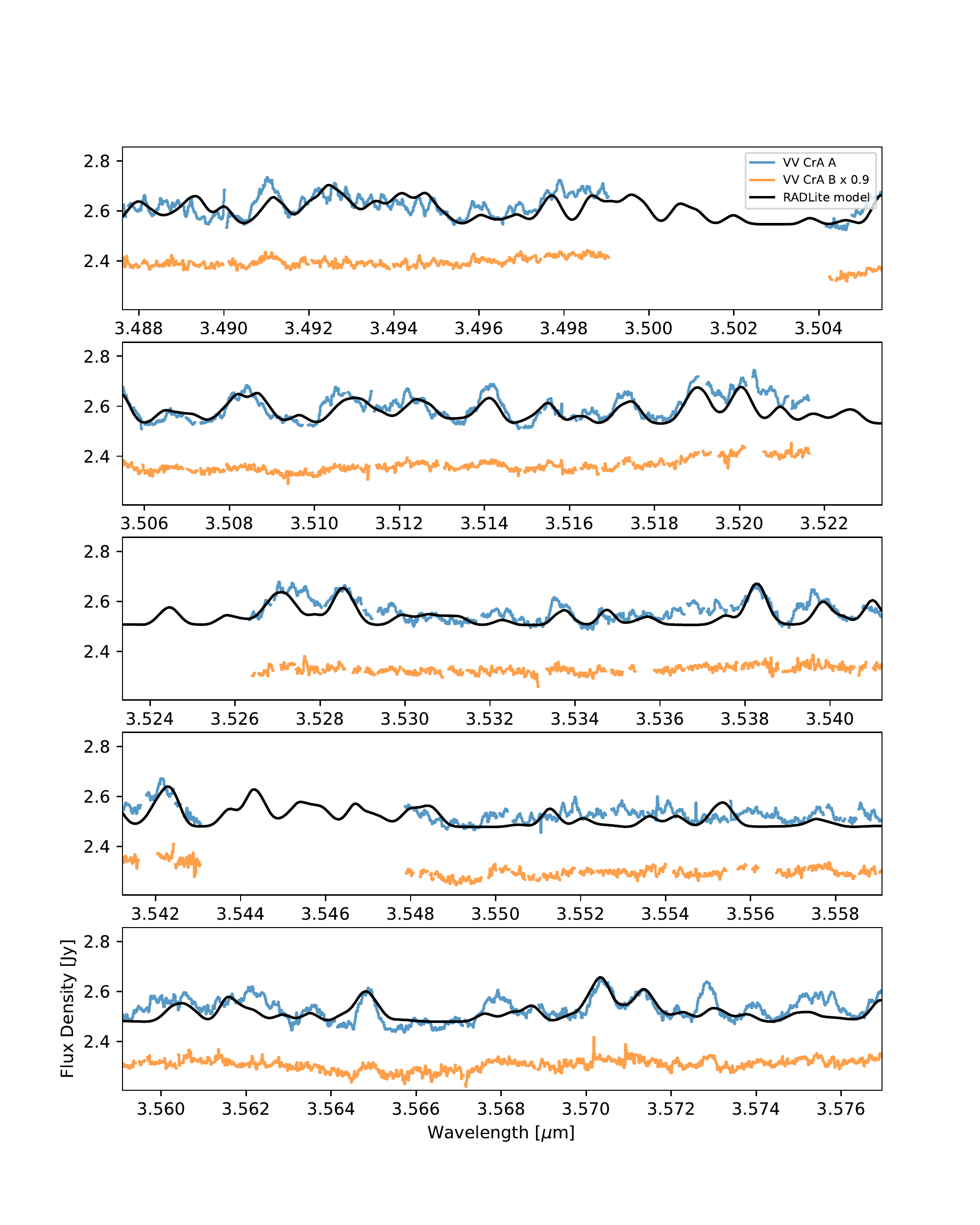}
\caption{(b) Second part of reduced CRIRES spectra of VV CrA A amd B in the L-band ``organics'' setting.}
\end{figure*}

\begin{figure*}\ContinuedFloat
  \centering 
  \epsscale{1}
\plotone{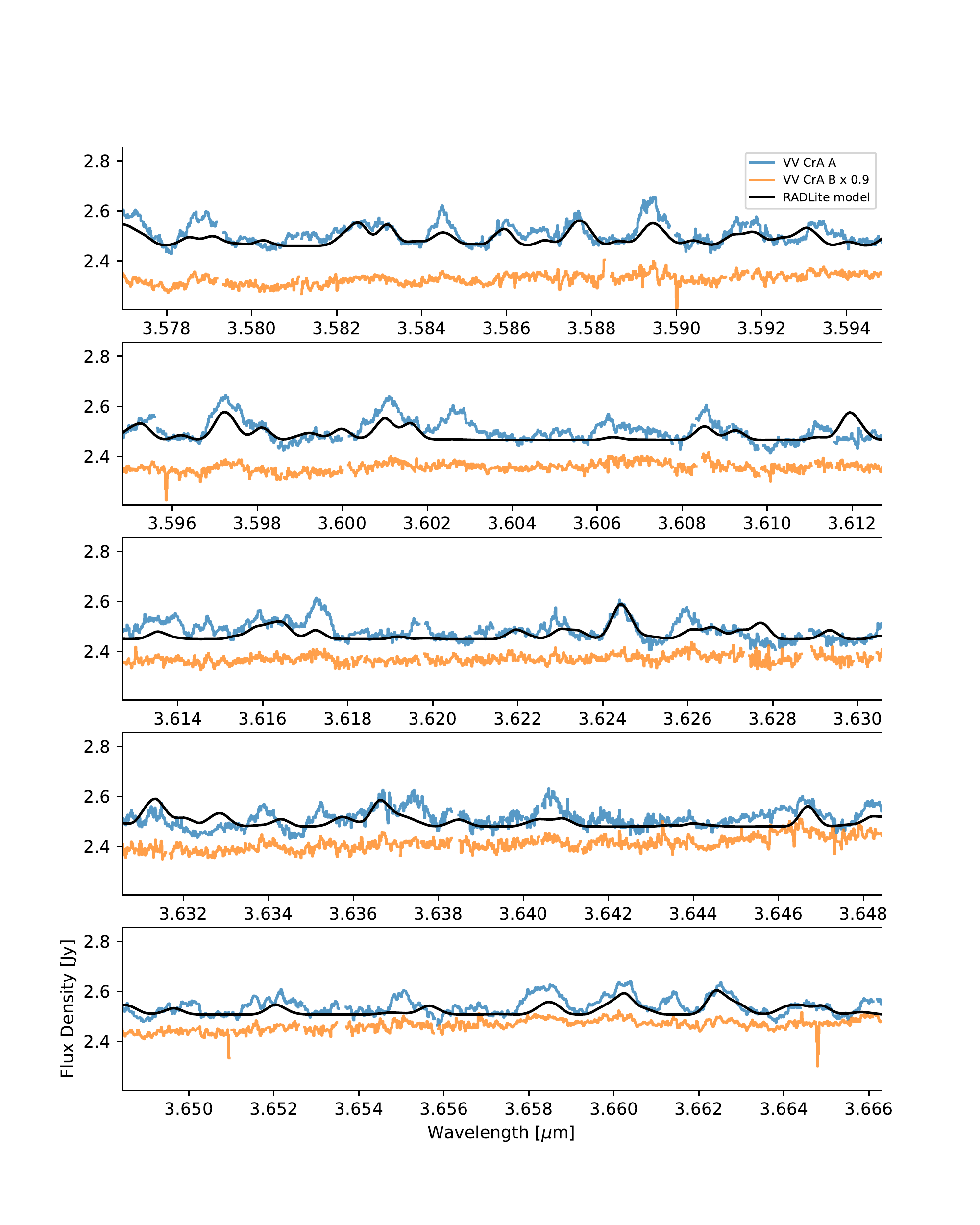}
\caption{(c) Third part of reduced CRIRES spectra of VV CrA A and B in the L-band ``organics'' setting.}
\end{figure*}

\begin{figure*}\ContinuedFloat
  \centering 
  \epsscale{1}
\plotone{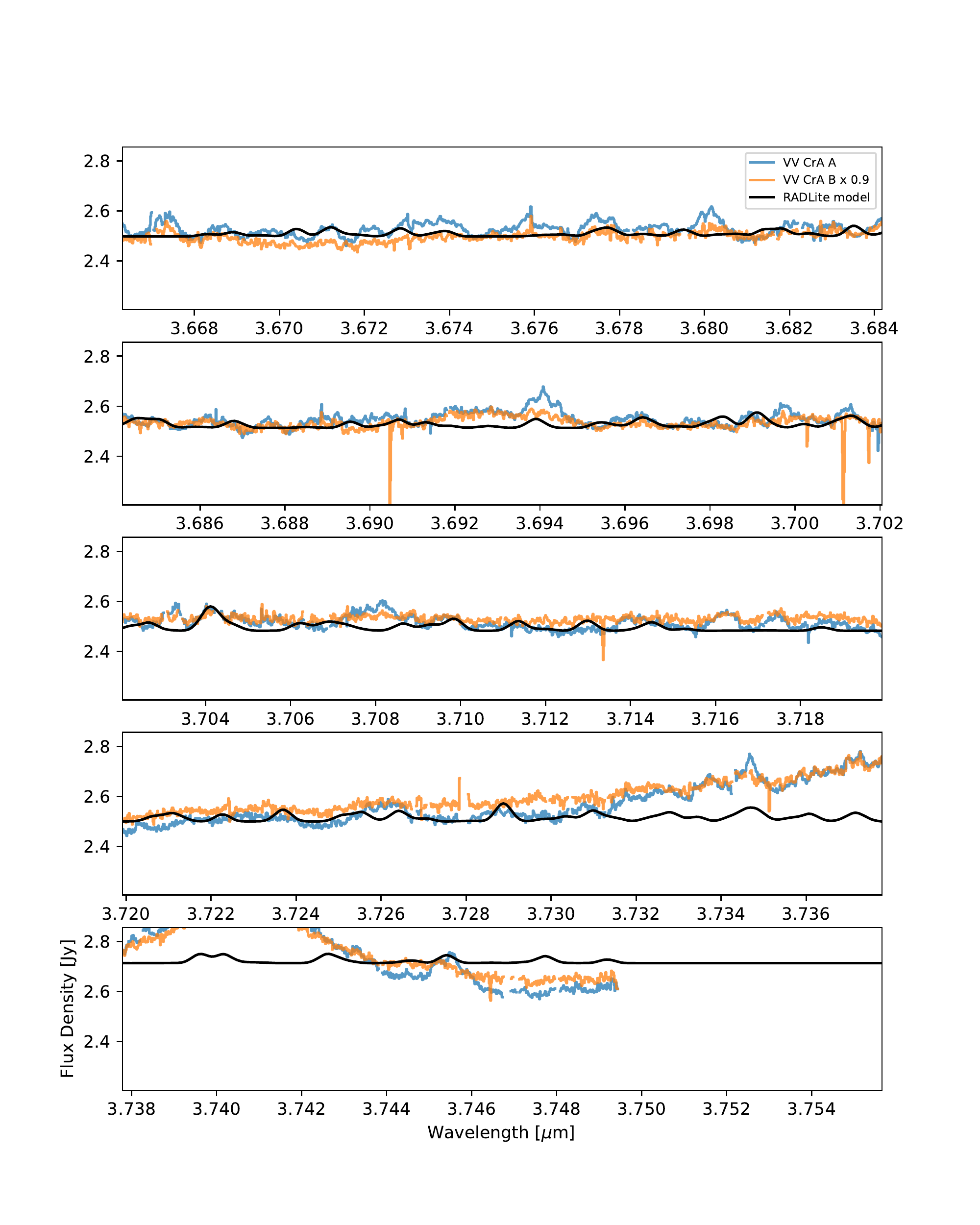}
\caption{(d) Fourth part of reduced CRIRES spectra of VV CrA A and B in the L-band ``organics'' setting.  The broad emission line at 3.741 $\mu$m is H I Pf$\gamma$.}
\end{figure*}

\end{document}